\batchmode
\makeatletter
\def\input@path{{"/home/jacob/Documents/Work/My Papers/2023-Stochastic Processes and Quantum Theory/"}}
\makeatother
\documentclass[11pt,twoside,english]{article}
\usepackage[T1]{fontenc}
\usepackage[latin9]{inputenc}
\usepackage{babel}
\usepackage{mathtools}
\usepackage{amsmath}
\usepackage{amssymb}
\usepackage{geometry}
\usepackage{setspace}
\usepackage{xargs}[2008/03/08]
\usepackage[pdftex,pdfusetitle,
 bookmarks=false,
 breaklinks=false,pdfborder={0 0 0},pdfborderstyle={},backref=false,colorlinks=false]
 {hyperref}

\makeatletter

\usepackage{endnotes}

\let\originalleft\left
\let\originalright\right
\renewcommand{\left}{\mathopen{}\mathclose\bgroup\originalleft}
\renewcommand{\right}{\aftergroup\egroup\originalright}

\def\smalloverbrace#1{\mathop{\vbox{\m@th\ialign{##\crcr%
      \noalign{\kern3\p@}%
      \tiny\downbracefill\crcr\noalign{\kern3\p@\nointerlineskip}%
      $\hfil\displaystyle{#1}\hfil$\crcr}}}\limits}

\def\smallunderbrace#1{\mathop{\vtop{\m@th\ialign{##\crcr
   $\hfil\displaystyle{#1}\hfil$\crcr
   \noalign{\kern3\p@\nointerlineskip}%
   \tiny\upbracefill\crcr\noalign{\kern3\p@}}}}\limits}

\DeclareMathAlphabet{\mymathbb}{U}{bbold}{m}{n}



\makeatother

\begin{document}
\title{The Stochastic-Quantum Theorem}
\author{Jacob A. Barandes\thanks{Departments of Philosophy and Physics, Harvard University, Cambridge, MA 02138; jacob\_barandes@harvard.edu; ORCID: 0000-0002-3740-4418}
}
\date{\today}

\maketitle

\begin{abstract}
This paper introduces several new classes of mathematical structures
that have close connections with physics and with the theory of dynamical
systems. The most general of these structures, called indivisible
stochastic processes, collectively encompass many important kinds
of stochastic processes, including Markov chains and random dynamical
systems. This paper then states and proves a new theorem that establishes
a precise correspondence between any indivisible stochastic process
and a unitarily evolving quantum system. This theorem therefore leads
to a new formulation of quantum theory, alongside the Hilbert-space,
path-integral, and quasi-probability formulations. The theorem also
provides a first-principles explanation for why quantum systems are
based on the complex numbers, Hilbert spaces, linear-unitary time
evolution, and the Born rule. In addition, the theorem suggests that
by selecting a suitable Hilbert space, together with an appropriate
choice of unitary evolution, one can simulate any indivisible stochastic
process on a quantum computer, thereby potentially opening up an extensive
set of novel applications for quantum computing.
\end{abstract}

\begin{center}
\global\long\def\quote#1{``#1"}%
\global\long\def\apostrophe{\textrm{'}}%
\global\long\def\slot{\phantom{x}}%
\global\long\def\eval#1{\left.#1\right\vert }%
\global\long\def\keyeq#1{\boxed{#1}}%
\global\long\def\importanteq#1{\boxed{\boxed{#1}}}%
\global\long\def\given{\vert}%
\global\long\def\mapping#1#2#3{#1:#2\to#3}%
\global\long\def\composition{\circ}%
\global\long\def\set#1{\left\{  #1\right\}  }%
\global\long\def\setindexed#1#2{\left\{  #1\right\}  _{#2}}%

\global\long\def\setbuild#1#2{\left\{  \left.\!#1\,\right|\,#2\right\}  }%
\global\long\def\suchthat{\,\vert\,}%
\global\long\def\complem{\mathrm{c}}%

\global\long\def\union{\cup}%
\global\long\def\intersection{\cap}%
\global\long\def\cartesianprod{\times}%
\global\long\def\disjointunion{\sqcup}%

\global\long\def\isomorphic{\cong}%

\global\long\def\setsize#1{\left|#1\right|}%
\global\long\def\defeq{\equiv}%
\global\long\def\conj{\ast}%
\global\long\def\overconj#1{\overline{#1}}%
\global\long\def\re{\mathrm{Re\,}}%
\global\long\def\im{\mathrm{Im\,}}%

\global\long\def\transp{\mathrm{T}}%
\global\long\def\tr{\mathrm{tr}}%
\global\long\def\adj{\dagger}%
\global\long\def\diag#1{\mathrm{diag}\left(#1\right)}%
\global\long\def\dotprod{\cdot}%
\global\long\def\crossprod{\times}%
\global\long\def\Probability#1{\mathrm{Prob}\left(#1\right)}%
\global\long\def\Amplitude#1{\mathrm{Amp}\left(#1\right)}%
\global\long\def\cov{\mathrm{cov}}%
\global\long\def\corr{\mathrm{corr}}%

\global\long\def\absval#1{\left\vert #1\right\vert }%
\global\long\def\expectval#1{\left\langle #1\right\rangle }%
\global\long\def\op#1{\hat{#1}}%

\global\long\def\bra#1{\left\langle #1\right|}%
\global\long\def\ket#1{\left|#1\right\rangle }%
\global\long\def\braket#1#2{\left\langle \left.\!#1\right|#2\right\rangle }%

\global\long\def\parens#1{(#1)}%
\global\long\def\bigparens#1{\big(#1\big)}%
\global\long\def\Bigparens#1{\Big(#1\Big)}%
\global\long\def\biggparens#1{\bigg(#1\bigg)}%
\global\long\def\Biggparens#1{\Bigg(#1\Bigg)}%
\global\long\def\bracks#1{[#1]}%
\global\long\def\bigbracks#1{\big[#1\big]}%
\global\long\def\Bigbracks#1{\Big[#1\Big]}%
\global\long\def\biggbracks#1{\bigg[#1\bigg]}%
\global\long\def\Biggbracks#1{\Bigg[#1\Bigg]}%
\global\long\def\curlies#1{\{#1\}}%
\global\long\def\bigcurlies#1{\big\{#1\big\}}%
\global\long\def\Bigcurlies#1{\Big\{#1\Big\}}%
\global\long\def\biggcurlies#1{\bigg\{#1\bigg\}}%
\global\long\def\Biggcurlies#1{\Bigg\{#1\Bigg\}}%
\global\long\def\verts#1{\vert#1\vert}%
\global\long\def\bigverts#1{\big\vert#1\big\vert}%
\global\long\def\Bigverts#1{\Big\vert#1\Big\vert}%
\global\long\def\biggverts#1{\bigg\vert#1\bigg\vert}%
\global\long\def\Biggverts#1{\Bigg\vert#1\Bigg\vert}%
\global\long\def\Verts#1{\Vert#1\Vert}%
\global\long\def\bigVerts#1{\big\Vert#1\big\Vert}%
\global\long\def\BigVerts#1{\Big\Vert#1\Big\Vert}%
\global\long\def\biggVerts#1{\bigg\Vert#1\bigg\Vert}%
\global\long\def\BiggVerts#1{\Bigg\Vert#1\Bigg\Vert}%
\global\long\def\ket#1{\vert#1\rangle}%
\global\long\def\bigket#1{\big\vert#1\big\rangle}%
\global\long\def\Bigket#1{\Big\vert#1\Big\rangle}%
\global\long\def\biggket#1{\bigg\vert#1\bigg\rangle}%
\global\long\def\Biggket#1{\Bigg\vert#1\Bigg\rangle}%
\global\long\def\bra#1{\langle#1\vert}%
\global\long\def\bigbra#1{\big\langle#1\big\vert}%
\global\long\def\Bigbra#1{\Big\langle#1\Big\vert}%
\global\long\def\biggbra#1{\bigg\langle#1\bigg\vert}%
\global\long\def\Biggbra#1{\Bigg\langle#1\Bigg\vert}%
\global\long\def\braket#1#2{\langle#1\vert#2\rangle}%
\global\long\def\bigbraket#1#2{\big\langle#1\big\vert#2\big\rangle}%
\global\long\def\Bigbraket#1#2{\Big\langle#1\Big\vert#2\Big\rangle}%
\global\long\def\biggbraket#1#2{\bigg\langle#1\bigg\vert#2\bigg\rangle}%
\global\long\def\Biggbraket#1#2{\Bigg\langle#1\Bigg\vert#2\Bigg\rangle}%
\global\long\def\angs#1{\langle#1\rangle}%
\global\long\def\bigangs#1{\big\langle#1\big\rangle}%
\global\long\def\Bigangs#1{\Big\langle#1\Big\rangle}%
\global\long\def\biggangs#1{\bigg\langle#1\bigg\rangle}%
\global\long\def\Biggangs#1{\Bigg\langle#1\Bigg\rangle}%

\global\long\def\vec#1{\mathbf{#1}}%
\global\long\def\vecgreek#1{\boldsymbol{#1}}%
\global\long\def\idmatrix{\mymathbb{1}}%
\global\long\def\projector{P}%
\global\long\def\permutationmatrix{\Sigma}%
\global\long\def\densitymatrix{\rho}%
\global\long\def\krausmatrix{K}%
\global\long\def\stochasticmatrix{\Gamma}%
\global\long\def\lindbladmatrix{L}%
\global\long\def\dynop{\Theta}%
\global\long\def\timeevop{U}%
\global\long\def\hadamardprod{\odot}%
\global\long\def\tensorprod{\otimes}%

\global\long\def\inprod#1#2{\left\langle #1,#2\right\rangle }%
\global\long\def\normket#1{\left\Vert #1\right\Vert }%
\global\long\def\hilbspace{\mathcal{H}}%
\global\long\def\samplespace{\Omega}%
\global\long\def\configspace{\mathcal{C}}%
\global\long\def\statespace{\mathcal{X}}%
\global\long\def\phasespace{\mathcal{P}}%
\global\long\def\spectrum{\sigma}%
\global\long\def\restrict#1#2{\left.#1\right\vert _{#2}}%
\global\long\def\from{\leftarrow}%
\global\long\def\statemap{\omega}%
\global\long\def\degangle#1{#1^{\circ}}%
\global\long\def\trivialvector{\tilde{v}}%
\global\long\def\eqsbrace#1{\left.#1\qquad\right\}  }%
\newcommandx\cptpmap[2][usedefault, addprefix=\global, 1=, 2=]{\mathcal{E}_{#1}^{#2}}%
\global\long\def\setoftimes{\mathcal{T}}%
\global\long\def\rvalgebra{\mathcal{A}}%
\par\end{center}

\section{Introduction\label{sec:Introduction}}

In the development of physical theories, it sometimes turns out that
existing definitions are too conceptually limiting, and that more
flexible definitions are needed. Working with more flexible definitions
at a higher level of abstraction or generality may make it easier
to discover new connections or prove new theorems, which would then
also apply down at the lower level of the original definitions.

This paper will argue that by appropriately generalizing standard
definitions of dynamical systems to include various forms of non-Markovianity,
one can obtain novel classes of mathematical structures that encompass
an extensive array of physically important models. As with a traditionally
defined dynamical system, each such mathematical structure describes
a physical system moving deterministically or stochastically along
some trajectory in a configuration space, albeit with a more general
set of laws than according to standard definitions.\footnote{For pedagogical treatments of the standard theory of dynamical systems,
see, for instance, Devaney (1989); Strogatz (1994); or Katok, Hasselblatt
(1995)\nocite{Devaney:1989aitcds,Strogatz:1994ndac,KatokHasselblatt:1995ittmtods}.}

This paper also states and proves a new theorem showing that despite
being based on trajectories in configuration spaces, the newly introduced
class of indivisible stochastic processes actually includes all quantum
systems with finite-dimensional Hilbert spaces.\footnote{It is worth keeping in mind that ``finite'' can still mean \emph{very
large}, with a level of discreteness that can be well below any conceivable
experimental resolution.} As a consequence, this stochastic-quantum theorem potentially offers
a more  conceptually transparent way to understand quantum systems,
with superpositions no longer regarded as literal blends of physical
states. The theorem also provides a first-principles explanation for
features of quantum theory that are usually taken to be axiomatic,
including Hilbert spaces over the complex numbers, linear-unitary
evolution, and the Born rule.

Seen from another point of view, this stochastic-quantum correspondence
yields an alternative way to formulate quantum theory\textemdash a
formulation that is phrased in the language of trajectories unfolding
stochastically in configuration spaces according to ordinary notions
of probability. This alternative formulation is distinct from the
traditional Hilbert-space formulation (Dirac 1930, von Neumann 1932)\nocite{Dirac:1930pofm,vonNeumann:1932mgdq},
the path-integral formulation (Dirac 1933; Feynman 1942, 1948)\nocite{Dirac:1933tliqm,Feynman:1942tpolaiqm,Feynman:1948statnrqm},
and the quasi-probability formulation (Wigner 1932, Moyal 1949)\nocite{Wigner:1932otqcfte,Moyal:1949qmaast}.

From a more practical perspective, turning this stochastic-quantum
correspondence around suggests that unitarily evolving quantum systems
can be put to work simulating a very broad class of non-Markovian
stochastic processes, thereby potentially opening up an extensive
suite of new applications for quantum computers.

As this paper will explain, indivisible stochastic processes are inherently
non-Markovian. The vast majority of research on non-Markovianity in
quantum theory has focused on the phenomenological appearance of non-Markovian-like
time evolution for density matrices of \emph{open} quantum systems,
due to interactions with their environments and associated feedback
effects. 

A remarkable and important exception was the work of Glick and Adami
(2020)\nocite{GlickAdami:2020manmqm}, who constructed an iterated
generalization of the \textquoteleft Wigner's friend\textquoteright{}
thought experiment, whose original incarnation first showed up in
Everett's unpublished 1956 dissertation and was developed further
by Wigner several years later (Everett 1956, Wigner 1961)\nocite{Everett:1956ttotuwf,Wigner:1961rotmbq}.
In their paper, Glick and Adami analyzed a hypothetical collection
of small devices carrying out sequential measurements on a given quantum
system, all inside a perfectly sealed container. In particular, Glick
and Adami showed that it makes an \emph{observable difference} whether
those sequential measurements are treated as collapse events or not.
If the individual measurements are treated as collapse events, then
the overall system behaves effectively as a Markov chain, according
to a definition of Markovianity that coincides with the notion of
divisibility used in the present work. By contrast, if the individual
measurements are treated as a form of unitary time evolution, so that
the overall set-up is truly regarded as a \emph{closed} system, then
careful quantum-state tomography could reveal distinct empirical signatures
of non-Markovianity, meaning that a fundamental form of non-Markovianity
for closed systems has been lurking in quantum theory all along.

Section~\ref{sec:Deterministic-Systems} begins by defining deterministic
generalizations of dynamical systems, followed by the introduction
of important distinctions between indivisible, Markovian, and Markovian-homogeneous
dynamics. Section~\ref{sec:Stochastic-Processes} provides a generalized
definition of a system with stochastic laws, shows how to represent
such a system in the formalism of linear algebra, describes connections
between this work and the existing research literature, defines the
relationship between a composite system and its subsystems, and introduces
the crucial notion of a unistochastic process. Section~\ref{sec:The-Stochastic-Quantum Theorem}
states the stochastic-quantum theorem, whose proof is this paper's
primary goal, and then discusses some important corollaries and provides
a simple example of the theorem in practice. Section~\ref{sec:Proof-of-the-Theorem}
lays out the theorem's proof, which entails explicitly constructing
the claimed correspondence between stochastic processes and quantum
systems along the way. Section~\ref{sec:Discussion-and-Future-Work}
concludes the paper with a brief discussion of future work.

\section{Deterministic Systems\label{sec:Deterministic-Systems}}

\subsection{Indivisible Dynamical Systems\label{subsec:Indivisible-Dynamical-Systems}}

Dynamical systems are abstract mathematical structures that usefully
model many deterministic physical processes. According to the standard
definition (Devaney 1989; Strogatz 1994; Katok, Hasselblatt 1995)\nocite{Devaney:1989aitcds,Strogatz:1994ndac,KatokHasselblatt:1995ittmtods},
a dynamical system consists of a map representing some kind of evolution
law that can be applied repeatedly to the elements of some set of
states. A dynamical system is usually assumed to be divisible, in
the sense that one can \textquoteleft divide up\textquoteright{} its
evolution law over any time duration into well-defined evolution laws
that describe intermediate time durations.\footnote{Note that this terminology is unrelated to the much older concept
of \emph{infinite divisibility}, which refers to a probability distribution
that can be expressed as the probability distribution of a sum of
any integer number of independent and identically distributed random
variables.} The more general case would be an indivisible dynamical system that
might lack this feature.

The terms \textquoteleft divisible\textquoteright{} and \textquoteleft indivisible\textquoteright{}
for dynamical laws are remarkably new. This terminology appears to
be due to Wolf and Cirac, who introduced it in a 2008 paper on quantum
channels (Wolf, Cirac 2008)\nocite{WolfCirac:2008dqc}.

To accommodate the eventual possibility of indivisible evolution,
this paper will define an indivisible dynamical system to mean a tuple
of the form 
\begin{equation}
\left(\statespace,\setoftimes,f\right)\label{eq:DefIndivisibleDynamicalSystem}
\end{equation}
 that consists of the following data.
\begin{itemize}
\item The symbol $\statespace$ denotes a set that will be called the indivisible
dynamical system's state space (or phase space), and whose individual
elements $i\in\statespace$ denote the system's (allowed) states.
\item Note that $\statespace$ may or may not be a finite set, and it may
or may not involve additional structure in its definition, such as
a measure-theoretic structure or a vector-space structure. For the
purposes of this paper, no such additional structure will be specified
or assumed. More broadly, for reasons of brevity and simplicity, this
paper will entirely set aside measure-theoretic considerations that
arise for the case of uncountable sets.
\item The symbol $\setoftimes$ denotes the system's set of target times
$t\in\setoftimes$, where $\setoftimes$ may or may not be isomorphic
to a subset of the real line $\mathbb{R}$ under addition.
\item The symbol $f$ denotes a map
\begin{equation}
\mapping f{\statespace\cartesianprod\setoftimes}{\statespace}\label{eq:IndivisibleDynamicalSystemDefDynamicalMap}
\end{equation}
 that will be called the system's dynamical map. This dynamical map
$f$ takes as inputs any state $i$ and any target time $t$, and
outputs a state $f\left(i,t\right)\in\statespace$: 
\begin{equation}
i,t\mapsto f\left(i,t\right)\in\statespace\quad\left[\textrm{for all }i\in\statespace,\,t\in\setoftimes\right].\label{eq:IndivisibleDynamicalSystemDynamicalMapValues}
\end{equation}
\item Fixing the target time $t$ turns $f$ into a time-dependent dynamical
map 
\begin{equation}
\mapping{f_{t}}{\statespace}{\statespace}\label{eq:IndivisibleDynamicalSystemTimeDepDynamicalMap}
\end{equation}
 defined by 
\begin{equation}
i\mapsto f_{t}\left(i\right)\defeq f\left(i,t\right)\quad\left[\textrm{for all }i\in\statespace\right].\label{eq:IndivisibleDynamicalSystemTimeDepDynamicalMapValues}
\end{equation}
\item Without any important loss of generality, the set of target times
$\setoftimes$ will be assumed to include an element denoted by $0$
and called the initial time. It will be further assumed that at the
initial time $0$, the time-dependent dynamical map $f_{t}$ trivializes
to the identity map $\mathrm{id}_{\statespace}$ on $\statespace$:
\begin{equation}
f_{0}=\mathrm{id}_{\statespace},\quad\textrm{or}\quad f_{0}\left(i\right)=i\quad\left[\textrm{for all }i\in\statespace\right].\label{eq:IndivisibleDynamicalSystemDynamicalMapInitialTimeTrivial}
\end{equation}
\item One can regard the argument $i$ appearing in the expression $f_{t}\left(i\right)$
as an initial state of the system at the initial time $0$, with the
time-dependent dynamical map $f_{t}$ then describing the evolution
of that state $i$ from the initial time $0$ to the target time $t$.
\item Given a fixed state $i$, the set of states 
\begin{equation}
\curlies{f_{t}\left(i\right)\suchthat t\in\setoftimes}\subset\statespace\label{eq:IndivisibleDynamicalSystemDefOrbitTrajectory}
\end{equation}
 describes the orbit, or trajectory, of the initial state $i$ through
the system's state space $\statespace$ according to the dynamical
map $f$.
\end{itemize}

In many applications, one takes the set of target times $\setoftimes$
to be a semigroup, meaning that the definition of $\setoftimes$ includes
an associative binary operation $\star$ (which is often denoted instead
by $+$ in the commutative case): 
\begin{equation}
t,t^{\prime}\mapsto t\star t^{\prime}\in\setoftimes\quad\left[\textrm{for all }t,t^{\prime}\in\setoftimes\right],\label{eq:DefBinarySemigroupOperationForSetOfTimes}
\end{equation}
\begin{equation}
\parens{t\star t^{\prime}}\star t^{\prime\prime}=t\star\parens{t^{\prime}\star t^{\prime\prime}}\quad\left[\textrm{for all }t,t^{\prime},t^{\prime\prime}\in\setoftimes\right].\label{eq:SetOfTimesAssociativeSemigroupOperation}
\end{equation}
 One usually also takes the initial time $0$ to be the identity element
under this binary operation, 
\begin{equation}
0\star t=t\star0=t\quad\left[\textrm{for all }t\in\setoftimes\right],\label{eq:InitialTimeAsIdentityElement}
\end{equation}
 in which case $\setoftimes$ becomes a monoid, meaning a semigroup
with an identity element. If, furthermore, every target time $t$
has an inverse $t^{\prime}$ such that $t\star t^{\prime}=t^{\prime}\star t=0$,
then $\setoftimes$ becomes a group.

\subsection{Markovian Dynamical Systems\label{subsec:Markovian-Dynamical-Systems}}

In the most general case, an indivisible dynamical system will not
provide a way to evolve a system from a \emph{non-initial} target
time $t^{\prime}\ne0$ to another target time $t$. An indivisible
dynamical system will also generically lack any means of \textquoteleft dividing
up\textquoteright{} the evolution from $0$ to $t\ne0$ into well-defined
forms of evolution over intermediate time durations between $0$ and
$t$ (even assuming that the set of times $\setoftimes$ has a notion
of ordering). To make contact with the kinds of dynamical systems
considered more widely in the research literature, it will therefore
be necessary to introduce a somewhat less general class of mathematical
structures.

This paper will define a Markovian (or divisible) dynamical system
to be a tuple of the form 
\begin{equation}
\left(\statespace,\setoftimes,g\right).\label{eq:DefDivisibleDynamicalSystem}
\end{equation}

\begin{itemize}
\item Here $\statespace$ is a state space and $\setoftimes$ is a set of
times (no longer called target times) forming a monoid, whose identity
element $0$ plays the role of an initial time, as usual.
\item The symbol $g$ denotes a map 
\begin{equation}
\mapping g{\statespace\cartesianprod\setoftimes^{2}}{\statespace}\label{eq:DivisibleDynamicalSystemDefTransitionMap}
\end{equation}
 that will be called the Markovian dynamical system's transition map.
This transition map $g$ takes as inputs any state $i$ and any pair
of times $t,t^{\prime}$, and outputs a state $g\left(i,\left(t,t^{\prime}\right)\right)\in\statespace$:
\begin{align}
i,t,t^{\prime} & \mapsto g\left(i,\left(t,t^{\prime}\right)\right)\in\statespace\label{eq:DivisibleDynamicalSystemDynamicalMapValues}\\
 & \quad\left[\textrm{for all }i\in\statespace,\,t,t^{\prime}\in\setoftimes\right].\nonumber 
\end{align}
The transition map $g$ here should be understood as describing the
evolution or transition of the state $i$ at the time $t^{\prime}$
to the state $g\left(i,\left(t,t^{\prime}\right)\right)$ at the time
$t$.
\item Fixing two times $t,t^{\prime}$ turns $g$ into a time-dependent
transition map 
\begin{equation}
\mapping{g_{t\from t^{\prime}}}{\statespace}{\statespace}\label{eq:DivisibleDynamicalSystemTimeDepTransitionMap}
\end{equation}
 defined by 
\begin{align}
i & \mapsto g_{t\from t^{\prime}}\left(i\right)\defeq g\left(i,\left(t,t^{\prime}\right)\right)\label{eq:DivisibleDynamicalSystemTimeDepTransitionMapValues}\\
 & \quad\left[\textrm{for all }i\in\statespace\right].\nonumber 
\end{align}
 This time-dependent transition map will be required to trivialize
to the identity map $\mathrm{id}_{\statespace}$ on $\statespace$
when $t^{\prime}=t$, 
\begin{align}
g_{t\from t} & =\mathrm{id}_{\statespace}\quad\left[\textrm{for all }t\in\setoftimes\right],\label{eq:DivisibleDynamicalSystemTransitionMapTrivialization}\\
 & \negthickspace\negthickspace\negthickspace\negthickspace\textrm{or}\quad g_{t\from t}\left(i\right)=i\quad\left[\textrm{for all }i\in\statespace,\,t\in\setoftimes\right],\nonumber 
\end{align}
 as well as satisfy the \textquoteleft divisibility condition\textquoteright{}
\begin{align}
g_{t\from t^{\prime\prime}} & =g_{t\from t^{\prime}}\composition g_{t^{\prime}\from t^{\prime\prime}}\label{eq:DivisibleDynamicalSystemDivisibilityCondition}\\
 & \negthickspace\negthickspace\negthickspace\left[\textrm{for all }t,t^{\prime},t^{\prime\prime}\in\setoftimes\right],\nonumber 
\end{align}
 where $\composition$ denotes function composition. The divisibility
condition means that the time-dependent transition map $g_{t\from t^{\prime\prime}}$
factorizes or \textquoteleft divides up\textquoteright{} into a part
$g_{t^{\prime}\from t^{\prime\prime}}$ that carries out the evolution
from $t^{\prime\prime}$ to $t^{\prime}$, followed by a part $g_{t\from t^{\prime}}$
that carries out the evolution from $t^{\prime}$ to $t$.
\item Choosing the time $t^{\prime}$ in the time-dependent transition map
$g_{t\from t^{\prime}}$ to be the initial time $0$ naturally defines
a time-dependent dynamical map \eqref{eq:IndivisibleDynamicalSystemTimeDepDynamicalMap},
\begin{equation}
f_{t}\defeq g_{t\from0}\quad\left[\textrm{for all }t\in\setoftimes\right],\label{eq:DivisibleDynamicalSystemDefTimeDepDynamicalMapFromTransitionMap}
\end{equation}
 which then also defines an overall dynamical map $\mapping f{\statespace\cartesianprod\setoftimes}{\statespace}$
according to $f\left(i,t\right)\defeq f_{t}\left(i\right)$. The trivialization
condition $g_{0\from0}=\mathrm{id}_{\statespace}$ from \eqref{eq:DivisibleDynamicalSystemTransitionMapTrivialization}
ensures that $f_{t}$ satisfies the corresponding trivialization condition
$f_{0}=\mathrm{id}_{\statespace}$ in \eqref{eq:IndivisibleDynamicalSystemDynamicalMapInitialTimeTrivial}.
It follows that every Markovian dynamical system is, in particular,
a special case of an indivisible dynamical system, one that includes
the additional structure that corresponds to having a transition map
$g$.
\item Meanwhile, setting $t^{\prime\prime}=0$ in the divisibility condition
\eqref{eq:DivisibleDynamicalSystemDivisibilityCondition} on the transition
map $g$ yields the subsidiary divisibility condition 
\begin{equation}
f_{t}=g_{t\from t^{\prime}}\composition f_{t^{\prime}}\quad\left[\textrm{for all }t,t^{\prime}\in\setoftimes\right],\label{eq:DivisibleDynamicalSystemDivisibilityConditionForDynamicalMap}
\end{equation}
which means that the time-dependent dynamical map $f_{t}$ \textquoteleft divides
up\textquoteright{} into a part $f_{t^{\prime}}$ that carries out
the evolution from the initial time $0$ to $t^{\prime}$, followed
by a part $g_{t\from t^{\prime}}$ that carries out the evolution
from $t^{\prime}$ to $t$.
\item It also follows from the subsidiary divisibility condition \eqref{eq:DivisibleDynamicalSystemDivisibilityConditionForDynamicalMap},
together with the trivialization condition $f_{0}=\mathrm{id}_{\statespace}$
in \eqref{eq:IndivisibleDynamicalSystemDynamicalMapInitialTimeTrivial},
that the time-dependent dynamical map $f_{t}$ has a well-defined
inverse $f_{t}^{-1}$ given by 
\begin{equation}
f_{t}^{-1}=g_{0\from t}\quad\left[\textrm{for all }t\in\setoftimes\right].\label{eq:DivisibleDynamicalSystemInverseDynamicalMap}
\end{equation}
That is, a Markovian dynamical system's time-dependent dynamical maps
$f_{t}$ are invertible.\footnote{If the set of times $\setoftimes$ has a well-defined ordering relation,
and if one modifies the definition of a Markovian dynamical system
$\left(\statespace,\setoftimes,g\right)$ by restricting the transition
map \eqref{eq:DivisibleDynamicalSystemDefTransitionMap} so that $g_{t\from t^{\prime}}$
is only defined when $t^{\prime}$ comes before $t$ according to
that ordering relation, then the arguments leading to \eqref{eq:DivisibleDynamicalSystemInverseDynamicalMap}
will break down. In that case, the time-dependent dynamical maps $f_{t}$
will not necessarily have inverses.}
\end{itemize}

\subsection{Markovian-Homogeneous Dynamical Systems\label{subsec:Markovian-Homogeneous-Dynamical-Systems}}

A Markovian dynamical system will be called \emph{Markovian-homogeneous}
(or \emph{time-homogeneous}) if it has the special property 
\begin{equation}
g_{t\star t^{\prime}\from t^{\prime}}=g_{t\from0}\quad\left[\textrm{for all }t,t^{\prime}\in\setoftimes\right].\label{eq:DivisibleDynamicalSystemHomogeneityCondition}
\end{equation}
 Equivalently, using $f_{t}\defeq g_{t\from0}$ from \eqref{eq:DivisibleDynamicalSystemDefTimeDepDynamicalMapFromTransitionMap},
one can write this special property as 
\begin{equation}
g_{t\star t^{\prime}\from t^{\prime}}=f_{t}\quad\left[\textrm{for all }t,t^{\prime}\in\setoftimes\right].\label{eq:DivisibleDynamicalSystemHomogeneityConditionFromDynamicalMap}
\end{equation}
 It then follows immediately from the subsidiary divisibility condition
\eqref{eq:DivisibleDynamicalSystemDivisibilityConditionForDynamicalMap}
that the dynamical map $\mapping f{\statespace\cartesianprod\setoftimes}{\statespace}$
gives a semigroup action of the set of times $\setoftimes$ on the
state space $\statespace$, in the sense that the semigroup operation
$\star$ on $\setoftimes$ is mapped to function composition: 
\begin{equation}
f_{t\star t^{\prime}}=f_{t}\composition f_{t^{\prime}}\quad\left[\textrm{for all }t,t^{\prime}\in\setoftimes\right].\label{eq:DynamicalSystemSemigroupActionHomogeneityProperty}
\end{equation}
 This equation is an example of homogeneity in time, which implies
that time evolution depends only on duration and not on absolute times.

Notice that the homogeneity property \eqref{eq:DynamicalSystemSemigroupActionHomogeneityProperty}
is phrased directly in terms of a dynamical map $f$, so it can be
imposed on an indivisible dynamical system $\left(\statespace,\setoftimes,f\right)$,
without any need to invoke a transition map $g$. In that case, it
also follows that there is no longer a meaningful distinction between
an indivisible dynamical system $\left(\statespace,\setoftimes,f\right)$
and a Markovian dynamical system $\left(\statespace,\setoftimes,g\right)$.
As such, if the homogeneity property \eqref{eq:DynamicalSystemSemigroupActionHomogeneityProperty}
is imposed, then the resulting mathematical structure will simply
be called a Markovian-homogeneous dynamical system.

What most references call a dynamical system corresponds to what this
paper would call a Markovian-homogeneous dynamical system. That is,
a dynamical system, without any further qualifiers, is a tuple $\left(\statespace,\setoftimes,f\right)$
for which $\statespace$ is a state space, $\setoftimes$ is a set
of times forming a monoid, and $\mapping f{\statespace\cartesianprod\setoftimes}{\statespace}$
is a dynamical map satisfying the homogeneity property \eqref{eq:DynamicalSystemSemigroupActionHomogeneityProperty}.
It follows that an indivisible dynamical system \eqref{eq:DefIndivisibleDynamicalSystem}
is a generalization of the kinds of dynamical systems that are usually
considered in textbooks, in the research literature, and in many applications.

In the most general cases, however, indivisible dynamical systems
will not satisfy the homogeneity property \eqref{eq:DynamicalSystemSemigroupActionHomogeneityProperty}.
Nor will it necessarily be possible to obtain a given indivisible
dynamical system $\left(\statespace,\setoftimes,f\right)$ by starting
with a Markovian dynamical system $\left(\statespace,\setoftimes,g\right)$
and then defining a dynamical map $f$ according to $f_{t}\defeq g_{t\from0}$
for all times $t$, as in \eqref{eq:DivisibleDynamicalSystemDefTimeDepDynamicalMapFromTransitionMap}.
Indeed, in the case in which there do not exist inverse time-dependent
dynamical maps $f_{t}^{-1}$, then, in light of the inversion formula
\eqref{eq:DivisibleDynamicalSystemInverseDynamicalMap}, there would
be an immediate obstruction to deriving a dynamical map $f$ from
a transition map $g$. 

An indivisible dynamical system $\left(\statespace,\setoftimes,f\right)$
that cannot be derived from a Markovian dynamical system $\left(\statespace,\setoftimes,g\right)$
will lack the structure needed to divide up its time evolution into
intermediate durations.

\section{Stochastic Processes\label{sec:Stochastic-Processes}}

\subsection{Indivisible Stochastic Processes\label{subsec:Indivisible-Stochastic-Processes}}

This paper will be concerned primarily with a mathematical structure
that replaces the deterministic behavior of an indivisible dynamical
system \eqref{eq:DefIndivisibleDynamicalSystem} with probabilistic,
or stochastic, behavior. An indivisible stochastic process  will
be defined to mean a tuple of the form 
\begin{equation}
\left(\configspace,\setoftimes,\setoftimes_{0},\stochasticmatrix,p,\rvalgebra\right)\label{eq:DefIndivisibleStochasticSystem}
\end{equation}
 that consists of the following data. 
\begin{itemize}
\item The symbol $\configspace$ denotes a set called the system's configuration
space, and the elements of $\configspace$ are called the system's
(allowed) configurations. Configurations and configuration spaces
will play an analogous role for indivisible stochastic processes that
states and state spaces play for indivisible dynamical systems.
\item The reason for switching the terminology from \textquoteleft states\textquoteright{}
to \textquoteleft configurations\textquoteright{} is conceptual. In
applications of the theory of dynamical systems to physical situations,
such as in classical Hamiltonian mechanics, the notion of a \textquoteleft state\textquoteright{}
is often taken to include rates of change or momenta in addition to
configurations, because defining a \textquoteleft state\textquoteright{}
in that way can make it possible to obtain deterministic laws in the
form of first-order differential equations. For an indivisible stochastic
process, by contrast, the probabilistic laws may mean that rates of
change and momenta are not well-defined in the absence of specifying
a trajectory. In that case, the only available notion of a \textquoteleft state\textquoteright{}
is limited to the more rudimentary notion of a \textquoteleft configuration,\textquoteright{}
which is more like a static arrangement of things.
\item For the purposes of this paper, $\configspace$ will be assumed to
be a finite set, with some (possibly very large) number $N$ of elements
denoted by $1,2,3,\dots,N$, and indexed by Latin letters $i,j,\dots$.
For brevity, and by an abuse of notation, the symbol $\configspace$
will sometimes be used to refer to the indivisible stochastic process
as a whole.
\item The symbol $\setoftimes$ denotes the system's set of target times,
including a time $0$ that will be called the system's initial time.
\item The symbol $\mathcal{T}_{0}$ denotes the system's set of conditioning
times, and is taken to be a subset $\setoftimes_{0}\subset\setoftimes$
of the set of target times. Without any real loss of generality, $\setoftimes_{0}$
will be assumed to include the initial time $0$. In practical cases,
$\setoftimes_{0}$ will often be assumed to be a \textquoteleft sparse\textquoteright{}
subset of $\setoftimes$, in the sense that $\setoftimes$ will contain
many times not in $\setoftimes_{0}$.
\item The symbol $\stochasticmatrix$ denotes a map 
\begin{equation}
\mapping{\stochasticmatrix}{\configspace^{2}\cartesianprod\setoftimes\cartesianprod\setoftimes_{0}}{\left[0,1\right]}\subset\mathbb{R}\label{eq:DefStochasticMap}
\end{equation}
 that will be called the transition map of the indivisible stochastic
process, where $\left[0,1\right]$ denotes the closed unit interval
$0\leq x\leq1$. Each value of this transition map will be labeled
as 
\begin{align}
\stochasticmatrix_{ij}\left(t\from t_{0}\right) & \defeq\stochasticmatrix\left(\left(i,j\right),\left(t,t_{0}\right)\right)\label{eq:IndividualTransitionProbabilities}\\
 & \negthickspace\negthickspace\negthickspace\left[\textrm{for all }i,j\in\configspace,\,t\in\setoftimes,\,t_{0}\in\setoftimes_{0}\right],\nonumber 
\end{align}
 and will be called the conditional transition probability $p\left(i,t\given j,t_{0}\right)$
for the system to be in its $i$th configuration at the target time
$t$, given that the system is in its $j$th configuration at the
conditioning time $t_{0}$: 
\begin{align}
\stochasticmatrix_{ij}\left(t\from t_{0}\right) & \defeq p\left(i,t\given j,t_{0}\right)\in\left[0,1\right]\label{eq:IndividualTransitionProbabilityAsConditionalProbability}\\
 & \negthickspace\negthickspace\negthickspace\left[\textrm{for all }i,j\in\configspace,\,t\in\setoftimes,\,t_{0}\in\setoftimes_{0}\right].\nonumber 
\end{align}
 
\item The transition map $\stochasticmatrix$ is required to satisfy the
standard normalization condition 
\begin{align}
\sum_{i=1}^{N}\stochasticmatrix_{ij}\left(t\from t_{0}\right) & =\sum_{i=1}^{N}p\left(i,t\given j,t_{0}\right)=1\label{eq:ConditionalProbabilityNormalization}\\
 & \negthickspace\negthickspace\negthickspace\left[\textrm{for all }j\in\configspace,\,t\in\setoftimes,\,t_{0}\in\setoftimes_{0}\right],\nonumber 
\end{align}
 as well as the trivialization condition 
\begin{align}
\stochasticmatrix_{ij}\left(t_{0}\from t_{0}\right) & \defeq\delta_{ij}\defeq\begin{cases}
1 & \textrm{for }i=j,\\
0 & \textrm{for }i\ne j
\end{cases}\label{eq:StochasticMapInitialCondition}\\
 & \negthickspace\negthickspace\negthickspace\left[\textrm{for all }t_{0}\in\setoftimes_{0}\right],\nonumber 
\end{align}
 where $\delta_{ij}$ is the usual Kronecker delta. 
\item The symbol $p$ denotes a map 
\begin{equation}
\mapping p{\configspace\cartesianprod\setoftimes}{\left[0,1\right]}\subset\mathbb{R}\label{eq:ProbabilityDistributionMap}
\end{equation}
 that will be called the system's standalone probability distribution.
Each value of this map will be labeled as 
\begin{align}
p_{i}\left(t\right) & \defeq p\left(i,t\right)\in\left[0,1\right]\label{eq:IndividualStandaloneProbabilities}\\
 & \negthickspace\negthickspace\negthickspace\left[\textrm{for all }i\in\configspace,\,t\in\setoftimes\right],\nonumber 
\end{align}
 and will be called the standalone probability for the system to be
in its $i$th configuration at the target time $t$. 
\item Only the standalone probabilities $p_{1}\left(0\right),\dots,p_{N}\left(0\right)$
at the initial time $0$ will be taken to be freely adjustable, subject
to the standard normalization condition 
\begin{equation}
\sum_{j=1}^{N}p_{j}\left(0\right)=1.\label{eq:InitialStandaloneProbabilitiesNormalization}
\end{equation}
 The standalone probabilities at every other target time $t$ will
be assumed to be defined by the following law of total probability
or marginalization condition: 
\begin{align}
p_{i}\left(t\right) & \defeq\sum_{j=1}^{N}\stochasticmatrix_{ij}\left(t\from0\right)p_{j}\left(0\right)\label{eq:ConditionalBayesianMarginalizationCondition}\\
 & =\sum_{j=1}^{N}p\left(i,t\given j,0\right)p_{j}\left(0\right)\nonumber \\
 & \negthickspace\negthickspace\negthickspace\left[\textrm{for all }i\in\configspace,\,t\in\setoftimes\right].\nonumber 
\end{align}
 The normalization condition \eqref{eq:ConditionalProbabilityNormalization}
on the transition map $\stochasticmatrix$ and the normalization condition
\eqref{eq:InitialStandaloneProbabilitiesNormalization} on the standalone
probabilities at the initial time $0$ then together ensure that the
probability distribution $p$ satisfies the standard normalization
condition more generally: 
\begin{equation}
\sum_{i=1}^{N}p_{i}\left(t\right)=1\quad\left[\textrm{for all }t\in\setoftimes\right].\label{eq:StandaloneProbabilitiesNormalization}
\end{equation}
 
\item Note that the definition of the transition map $\stochasticmatrix$
is independent of the choice of standalone probabilities $p_{1}\left(0\right),\dots,p_{N}\left(0\right)$
at the initial time $0$. That is, $\stochasticmatrix$ can be freely
adjusted independently of those initial standalone probabilities.
Importantly, the law of total probability \eqref{eq:ConditionalBayesianMarginalizationCondition}
therefore defines a \emph{linear relationship} between the standalone
probabilities $p_{1}\left(0\right),\dots,p_{N}\left(0\right)$ at
the initial time $0$ and the standalone probabilities $p_{1}\left(t\right),\dots,p_{N}\left(t\right)$
at any target time $t$. In the work ahead, it will be argued that
this linear relationship is ultimately responsible for the linearity
of time evolution in quantum theory.
\item The transition map $\stochasticmatrix$ will be assumed to satisfy
the following divisibility condition for any target time $t\in\setoftimes$
and any pair of conditioning times $t_{0},t^{\prime}\in\setoftimes_{0}\subset\setoftimes$:
\begin{align}
\stochasticmatrix_{ij}\left(t\from t_{0}\right) & =\sum_{k=1}^{N}\stochasticmatrix_{ik}\left(t\from t^{\prime}\right)\stochasticmatrix_{kj}\left(t^{\prime}\from t_{0}\right)\label{eq:DivisibilityCondition}\\
 & \negthickspace\negthickspace\negthickspace\left[\textrm{for all }i,j\in\configspace,\,t\in\setoftimes,\,t_{0},t^{\prime}\in\setoftimes_{0}\right].\nonumber 
\end{align}
 As such, conditioning times like $t^{\prime}$ will alternatively
be called division events.
\item If $t^{\prime}$ is a target time but \emph{not} a conditioning time,
then the values $\stochasticmatrix_{ik}\left(t\from t^{\prime}\right)$
appearing in the divisibility condition \eqref{eq:DivisibilityCondition}
will not be well-defined, and the divisibility condition will not
hold.\footnote{Note that if the quantities $\stochasticmatrix_{kj}\left(t^{\prime}\from t_{0}\right)$
are regarded as forming an $N\times N$ stochastic matrix, a term
to be defined shortly, then the inverse matrix, if it exists, will
always have negative entries unless both matrices are permutation
matrices, as follows from an elementary theorem of linear algebra.
It follows that one cannot safely define the non-negative quantities
$\stochasticmatrix_{ik}\left(t\from t^{\prime}\right)$ by invoking
inverse matrices.} The process described here is therefore indivisible for generic target
times $t^{\prime}$, in a sense that parallels the notion of indivisibility
for an indivisible dynamical system, as introduced earlier in this
paper.
\item By combining the marginalization condition \eqref{eq:ConditionalBayesianMarginalizationCondition}
with the divisibility condition \eqref{eq:DivisibilityCondition},
it follows that the standalone probabilities $p_{1}\left(t\right),\dots,p_{N}\left(t\right)$
at any target time $t\in\setoftimes$ and the standalone probabilities
$p_{1}\left(t^{\prime}\right),\dots,p_{N}\left(t^{\prime}\right)$
at any conditioning time $t^{\prime}$ are related by the following
marginalization condition of their own: 
\begin{align}
p_{i}\left(t\right) & =\sum_{j=1}^{N}\stochasticmatrix_{ij}\left(t\from t^{\prime}\right)p_{j}\left(t^{\prime}\right)\label{eq:ConditionalBayesianMarginalizationConditionForGenericConditioningTimes}\\
 & =\sum_{j=1}^{N}p\left(i,t\given j,t^{\prime}\right)p_{j}\left(t^{\prime}\right)\\
 & \negthickspace\negthickspace\negthickspace\left[\textrm{for all }i\in\configspace,\,t\in\setoftimes,\,t^{\prime}\in\setoftimes_{0}\right].
\end{align}
\item The transition map $\stochasticmatrix$ will \emph{not} be assumed
to satisfy anything like a homogeneity property. More precisely, no
assumption will be made that given any pair of target times $t,t^{\prime}$,
there will exist a target time $t^{\prime\prime}$ such that the following
homogeneity property holds: 
\begin{align}
\stochasticmatrix_{ij}\left(t\from0\right) & =\sum_{k=1}^{N}\stochasticmatrix_{ik}\left(t^{\prime\prime}\from0\right)\stochasticmatrix_{kj}\left(t^{\prime}\from0\right)\label{eq:HomogeneityPropertyForTimeDependentDynamicalMaps}\\
 & \negthickspace\negthickspace\negthickspace\left[\textrm{for all }i,j\in\configspace\right].\nonumber 
\end{align}
 That is, an indivisible stochastic process will generically be non-Markovian.
\item These notions of indivisibility and non-Markovianity represent a distinct
way to generalize the Markovian case, as compared with most forms
of non-Markovianity described in textbooks and in the research literature.
According to those more traditional forms of non-Markovianity, as
emphasized by Gillespie (1998, 2000)\nocite{Gillespie:1998dtsoasp,Gillespie:2000nsp},
one assumes that the set of target times $\setoftimes$ has an ordering
relation, and one further assumes the existence of higher-order conditional
probabilities $p\left(i,t\given j_{1},t_{1};j_{2},t_{2};\dots\right)$
that are conditioned on arbitrarily many conditioning times $t_{1},t_{2},\dots$.
From that more traditional standpoint, the system is Markovian if
and only if the latest conditioning time $t_{1}$ always screens off
all the earlier conditioning times $t_{2},\dots$, so that $p\left(i,t\given j_{1},t_{1};j_{2},t_{2};\dots\right)=p\left(i,t\given j_{1},t_{1}\right)$.
The definition of an indivisible stochastic process presented in this
paper does not assume fixed values of such higher-order conditional
probabilities in the first place, and therefore, in a sense, represents
a whole equivalence class of non-Markovian models, each of which is
called a non-Markovian realizer. In particular, an indivisible stochastic
process is defined by fixing less information than any of its non-Markovian
realizers.
\item The symbol $\rvalgebra$ denotes a commutative algebra of maps of
the form 
\begin{equation}
\mapping A{\configspace\cartesianprod\setoftimes}{\mathbb{R}}\label{eq:DefRandomVariable}
\end{equation}
 under the usual rules of function arithmetic, and will be called
the system's algebra of random variables.  The individual values
of each random variable $A\in\rvalgebra$ will be labeled as 
\begin{equation}
a_{i}\left(t\right)\defeq A\left(i,t\right)\in\mathbb{R}\quad\left[\textrm{for all }i\in\configspace,\,t\in\setoftimes\right].\label{eq:RandomVariableMagnitudes}
\end{equation}
 Each such value $a_{i}\left(t\right)$ will be called the magnitude
or value of the random variable $A$ when the system is in its $i$th
configuration at the target time $t$. No assumption will be made
here that these magnitudes are all distinct, even at any fixed target
time $t$.
\item For the purposes of this paper, the algebra of random variables $\rvalgebra$
will always be taken to be maximal, in the sense of containing every
well-defined map of the form \eqref{eq:DefRandomVariable}.
\item Fixing the target time $t$, the expectation value $\expectval{A\left(t\right)}$
of a random variable $A$ will denote its statistical average according
to the standalone probability distribution $p$ at the target time
$t$: 
\begin{equation}
\expectval{A\left(t\right)}\defeq\sum_{i=1}^{N}a_{i}\left(t\right)p_{i}\left(t\right).\label{eq:DefExpectationValueRandomVariable}
\end{equation}
\end{itemize}

\subsection{Ingredients from Linear Algebra\label{subsec:Ingredients-from-Linear-Algebra}}

Given an indivisible stochastic process $\left(\configspace,\setoftimes,\setoftimes_{0},\stochasticmatrix,p,\rvalgebra\right)$,
with $\configspace$ a configuration space of finite integer size
$N$, it will be convenient to introduce some formalism from linear
algebra.
\begin{itemize}
\item Fixing a target time $t$, let $p\left(t\right)$, called the system's
(time-dependent) probability vector, denote the $N\times1$ column
vector whose $i$th entry is the $i$th standalone probability $p_{i}\left(t\right)$:
\begin{equation}
p\left(t\right)\defeq\begin{pmatrix}p_{1}\left(t\right)\\
\vdots\\
p_{N}\left(t\right)
\end{pmatrix}.\label{eq:DefProbabilityVector}
\end{equation}
\item Fixing a target time $t$ and a conditioning time $t_{0}$, let $\stochasticmatrix\left(t\from t_{0}\right)$,
called the system's (time-dependent) transition matrix, denote the
$N\times N$ matrix for which the entry in the $i$th row, $j$th
column is the conditional transition probability $\stochasticmatrix_{ij}\left(t\from t_{0}\right)\defeq p\left(i,t\given j,t_{0}\right)$:
\begin{align}
 & \stochasticmatrix\left(t\from t_{0}\right)\defeq\begin{pmatrix}\stochasticmatrix_{11}\left(t\from t_{0}\right) & \stochasticmatrix_{12}\left(t\from t_{0}\right)\\
\stochasticmatrix_{21}\left(t\from t_{0}\right) & \ddots\\
 &  & \stochasticmatrix_{NN}\left(t\from t_{0}\right)
\end{pmatrix}\nonumber \\
 & =\begin{pmatrix}p\left(1,t\given1,t_{0}\right) & p\left(1,t\given2,t_{0}\right)\\
p\left(2,t\given1,t_{0}\right) & \ddots\\
 &  & p\left(N,t\given N,t_{0}\right)
\end{pmatrix}.\label{eq:DefTransitionMatrix}
\end{align}
\end{itemize}

It follows that for any target time $t$ and any conditioning time
$t_{0}$, the transition matrix $\stochasticmatrix\left(t\from t_{0}\right)$
is a (column) stochastic matrix in the mathematical sense, meaning
that its entries are all non-negative real numbers, 
\begin{equation}
\stochasticmatrix_{ij}\left(t\from t_{0}\right)\geq0\quad\left[\textrm{for all }i,j\in\configspace,\,t\in\setoftimes,\,t_{0}\in\setoftimes_{0}\right],\label{eq:TransitionProbabilitiesNonNegative}
\end{equation}
 and that its columns each sum to $1$, as required in \eqref{eq:ConditionalProbabilityNormalization}.
The trivialization condition \eqref{eq:StochasticMapInitialCondition}
on the transition map $\stochasticmatrix$ then becomes the statement
that the transition matrix $\stochasticmatrix\left(t_{0}\from t_{0}\right)$
for any conditioning time $t_{0}$ is just the $N\times N$ identity
matrix $\idmatrix$: 
\begin{equation}
\stochasticmatrix\left(t_{0}\from t_{0}\right)=\idmatrix\defeq\begin{pmatrix}1 & 0\\
0 & \ddots\\
 &  & 1
\end{pmatrix}.\label{eq:InitialConditionStochasticMatrix}
\end{equation}
Observe that the law of total probability \eqref{eq:ConditionalBayesianMarginalizationConditionForGenericConditioningTimes}
for any target time $t$ and any conditioning time $t^{\prime}$ naturally
takes the form of matrix multiplication: 
\begin{equation}
p\left(t\right)=\stochasticmatrix\left(t\from t^{\prime}\right)p\left(t^{\prime}\right)\quad\left[\textrm{for all }t\in\setoftimes,\,t^{\prime}\in\setoftimes_{0}\right].\label{eq:ConditionalMarginalizationConditionMatrixForm}
\end{equation}

\subsection{Connections with Other Constructions\label{subsec:Connections-with-Other-Constructions}}

There is a definite sense in which an indivisible stochastic process
$\left(\configspace,\setoftimes,\setoftimes_{0},\stochasticmatrix,p,\rvalgebra\right)$,
as introduced in \eqref{eq:DefIndivisibleStochasticSystem}, provides
a probabilistic extension of an indivisible dynamical system $\left(\statespace,\setoftimes,f\right)$,
as introduced in \eqref{eq:DefIndivisibleDynamicalSystem}. Indeed,
if the set of conditioning times $\setoftimes_{0}$ contains only
the initial time $0$, and the transition map $\mapping{\stochasticmatrix}{\configspace^{2}\cartesianprod\setoftimes\cartesianprod\setoftimes_{0}}{\left[0,1\right]}$
from \eqref{eq:DefStochasticMap} outputs only the trivial probabilities
$1$ and $0$, then $\stochasticmatrix$ is effectively deterministic.
In that case, one can naturally set $\statespace\defeq\configspace$
and define a dynamical map $\mapping f{\statespace\cartesianprod\setoftimes}{\statespace}$
according to 
\begin{align}
f\left(j,t\right) & =i\quad\textrm{if and only if}\quad\stochasticmatrix\left(\left(i,j\right),\left(t,0\right)\right)=1\label{eq:DynamicalMapDeterminedByTrivialStochasticMap}\\
 & \quad\left[\textrm{for all }i,j\in\statespace,\,t\in\setoftimes\right].\nonumber 
\end{align}

An indivisible stochastic process is distinct from a mathematical
structure already in the research literature, called a stochastic
dynamical system or a random dynamical system (Honerkamp 1996, Arnold
1998)\nocite{Honerkamp:1996sdscnmda,Arnold:1998rds}.
\begin{itemize}
\item The definition of a random dynamical system starts with what this
paper would call an indivisible dynamical system $\left(\statespace,\setoftimes,f\right)$,
with $\setoftimes$ a monoid, and replaces the single dynamical map
$f$ with a probabilistic family or ensemble of dynamical maps $\curlies{f_{\omega}\suchthat\omega\in\samplespace}$
that are indexed by a point $\omega$ in some sample space $\samplespace$.
That is, a randomly sampled point $\omega$ in $\samplespace$ picks
out an entire dynamical map $f_{\omega}$ as a whole.
\item For any randomly sampled point $\omega$ and any fixed target time
$t$, one can define a map 
\begin{equation}
\mapping{f_{\omega,t}}{\statespace}{\statespace}\label{eq:RandomDynamicalSystemDefTimeDepDynamicalMap}
\end{equation}
 by 
\begin{equation}
i\mapsto f_{\omega,t}\left(i\right)\defeq f_{\omega}\left(i,t\right)\quad\left[\textrm{for all }i\in\statespace\right].\label{eq:RandomDynamicalSystemDefTimeDepDynamicalMapValues}
\end{equation}
 A random dynamical system is then assumed to satisfy an initial condition
that generalizes \eqref{eq:IndivisibleDynamicalSystemDynamicalMapInitialTimeTrivial},
\begin{equation}
f_{\omega,0}=\mathrm{id}_{\statespace}\quad\left[\textrm{for all }\omega\in\samplespace\right],\label{eq:RandomDynamicalSystemDynamicalMapInitialTimeTrivial}
\end{equation}
 as well as a homogeneity property that generalizes \eqref{eq:DynamicalSystemSemigroupActionHomogeneityProperty},
\begin{align}
f_{\omega^{\prime},t\star t^{\prime}} & =f_{\theta\left(\omega^{\prime},t^{\prime}\right),t}\composition f_{\omega^{\prime},t^{\prime}}\label{eq:RandomDynamicalSystemMarkovProperty}\\
 & \negthickspace\negthickspace\negthickspace\left[\textrm{for all }\omega^{\prime}\in\samplespace,\,t,t^{\prime}\in\setoftimes\right].\nonumber 
\end{align}
 Here $\mapping{\theta}{\samplespace\cartesianprod\setoftimes}{\samplespace}$
is a map that is part of the definition of the random dynamical system,
and specifies a deterministic rule for updating the sample point $\omega^{\prime}$
and the dynamical map $f_{\omega^{\prime}}$ to accommodate replacing
the original initial time $0$ with the effectively new initial time
$t^{\prime}$. That is, $\theta$ is necessary to account for any
deterministic evolution in the underlying source of randomness itself.
\item It follows from the foregoing definitions that the conditional probability
$p\left(i,t\given j,0\right)$ for the system to be in its $i$th
state at the target time $t$, given that the system is in its $j$th
state at the initial time $0$, is obtained by adding up the probabilities
for all the points $\omega$ in the sample space $\samplespace$ whose
corresponding dynamical maps $f_{\omega}$ take the state $j$ at
the initial time $0$ and yield the state $i$ at the final time $t$:
\begin{align}
p\left(i,t\given j,0\right) & =\textrm{probability}\left(\curlies{\omega\in\Omega\suchthat f_{\omega,t}\left(j\right)=i}\right)\label{eq:ConditionalProbabilityFromRandomDynamicalSystem}\\
 & \negthickspace\negthickspace\negthickspace\left[\textrm{for all }i,j\in\statespace,\,t\in\setoftimes\right].\nonumber 
\end{align}
\end{itemize}

In a sense, the absolutely \emph{most} \emph{general} kind of stochastic
mathematical structure is simply called a stochastic process, and
essentially requires only the specification of a state space $\statespace$,
a set of target times $\setoftimes$, an initial probability distribution
$p$, and a set of one or more time-dependent random variables $\rvalgebra$.\footnote{See Rosenblatt (1962), Parzen (1962), Doob (1990), or Ross (1995)\nocite{Rosenblatt:1962rp,Parzen:1962sp,Doob:1990sp,Ross:1995sp}
for textbook treatments.} Importantly, the definition of a stochastic process lacks the specification
of a dynamical law. By requiring a dynamical law in the form of a
transition map $\stochasticmatrix$, an indivisible stochastic process
$\left(\configspace,\setoftimes,\setoftimes_{0},\stochasticmatrix,p,\rvalgebra\right)$
is not quite as general as a stochastic process, but will still be
general enough to encompass a large class of physical and mathematical
models.

In particular, one can regard any Markov chain as a special case of
an indivisible stochastic process. The starting place is to assume
that the set of target times $\setoftimes$ and the set of conditioning
times $\setoftimes_{0}$ are identical and are both isomorphic to
the integers, $\setoftimes=\setoftimes_{0}\cong\mathbb{Z}$, with
each time $t=n\,\delta t$ an integer number $n\in\mathbb{Z}$ of
steps of some fixed, elementary time scale $\delta t$. One then further
assumes that for each integer $n$, the time-dependent transition
matrix $\stochasticmatrix\left(n\,\delta t\from0\right)$ originally
defined in \eqref{eq:DefTransitionMatrix} can be expressed as the
$n$th power of the transition matrix $\stochasticmatrix\left(\delta t\right)\defeq\stochasticmatrix\left(\delta t\from0\right)$
that implements the evolution for just the first time step $\delta t$:
\begin{equation}
\stochasticmatrix\left(n\,\delta t\from0\right)=\left[\stochasticmatrix\left(\delta t\right)\right]^{n}\quad\left[\textrm{for all }n\in\mathbb{Z}\right].\label{eq:MarkovChain}
\end{equation}
More broadly, an indivisible stochastic process can therefore be
understood as a kind of non-Markovian generalization of a Markov chain.

As this paper will also show, the class of indivisible stochastic
processes essentially includes all quantum systems as well (at least
those that have or can be approximated as having finite-dimensional
Hilbert spaces).

\subsection{Composite Systems and Subsystems\label{subsec:Composite-Systems-and-Subsystems}}

Introducing a notation of primes and tildes now to distinguish between
 different indivisible stochastic processes, an indivisible stochastic
process $\parens{\tilde{\configspace},\tilde{\setoftimes},\tilde{\setoftimes}_{0},\tilde{\stochasticmatrix},\tilde{p},\tilde{\rvalgebra}}$
will be called a composite system if its configuration space $\tilde{\configspace}$
is naturally expressible as a nontrivial Cartesian product of two
sets $\configspace$ and $\configspace^{\prime}$: 
\begin{equation}
\tilde{\configspace}=\configspace\cartesianprod\configspace^{\prime}.\label{eq:CompositeSystemConfigSpaceCartesianProd}
\end{equation}
 A composite system has the following salient features.
\begin{itemize}
\item Letting $N$ denote the size of $\configspace$, and letting $N^{\prime}$
denote the size of $\configspace^{\prime}$, the composite system's
configuration space $\tilde{\configspace}$ has size $\tilde{N}\defeq NN^{\prime}$.
\item Labeling the elements of $\configspace$ by unprimed Latin letters
$i,j,\dots$, and labeling the elements of $\configspace^{\prime}$
by primed Latin letters $i^{\prime},j^{\prime},\dots$, the composite
system's transition map $\mapping{\tilde{\stochasticmatrix}}{\tilde{\configspace}^{2}\cartesianprod\tilde{\setoftimes}\cartesianprod\tilde{\setoftimes}_{0}}{\left[0,1\right]}$
has individual values that will be denoted by 
\begin{align}
\tilde{\stochasticmatrix}_{ii^{\prime},jj^{\prime}}\left(t\from t_{0}\right) & \defeq\tilde{\stochasticmatrix}\left(\left(\left(i,i^{\prime}\right),\left(j,j^{\prime}\right)\right),\left(t,t_{0}\right)\right)\nonumber \\
 & \defeq\tilde{p}\left(\left(i,i^{\prime}\right),t\given\left(j,j^{\prime}\right),t_{0}\right)\label{eq:DefTransitionProbabilitiesForCompositeSystem}\\
 & \negthickspace\negthickspace\negthickspace\bracks{\textrm{for all }i,j\in\configspace,\,i^{\prime},j^{\prime}\in\configspace^{\prime},\,t\in\tilde{\setoftimes},\,t_{0}\in\tilde{\setoftimes}_{0}}.\nonumber 
\end{align}
 The normalization condition \eqref{eq:ConditionalProbabilityNormalization}
on the transition map $\tilde{\stochasticmatrix}$ takes the form
\begin{align}
 & \sum_{i=1}^{N}\sum_{i^{\prime}=1}^{N^{\prime}}\tilde{\stochasticmatrix}_{ii^{\prime},jj^{\prime}}\left(t\from t_{0}\right)=1\label{eq:CompositeSystemConditionalProbabilityNormalization}\\
 & \quad\bracks{\textrm{for all }j\in\configspace,\,j^{\prime}\in\configspace^{\prime},\,t\in\tilde{\setoftimes},\,t_{0}\in\tilde{\setoftimes}_{0}}.\nonumber 
\end{align}
\item The composite system's probability distribution $\mapping{\tilde{p}}{\tilde{\configspace}\cartesianprod\tilde{\setoftimes}}{\left[0,1\right]}$
has individual values at any fixed target time $t$ that will be denoted
by 
\begin{align}
\tilde{p}_{ii^{\prime}}\left(t\right) & \defeq p\left(\left(i,i^{\prime}\right),t\right)\label{eq:DefJointProbabilitiesForCompositeSystem}\\
 & \negthickspace\negthickspace\negthickspace\bracks{\textrm{for all }i\in\configspace,\,i^{\prime}\in\configspace^{\prime},\,t\in\tilde{\setoftimes}}.\nonumber 
\end{align}
 The normalization condition \eqref{eq:StandaloneProbabilitiesNormalization}
on the probability distribution $\tilde{p}$ takes the form 
\begin{equation}
\sum_{i=1}^{N}\sum_{i^{\prime}=1}^{N^{\prime}}\tilde{p}_{ii^{\prime}}\left(t\right)=1\quad\bracks{\textrm{for all }t\in\tilde{\setoftimes}},\label{eq:JointProbabilitiesNormalizationCondition}
\end{equation}
 and the law of total probability \eqref{eq:ConditionalBayesianMarginalizationConditionForGenericConditioningTimes}
becomes 
\begin{align}
\tilde{p}_{ii^{\prime}}\left(t\right) & \defeq\sum_{j=1}^{N}\sum_{j^{\prime}=1}^{N^{\prime}}\tilde{\stochasticmatrix}_{ii^{\prime},jj^{\prime}}\left(t\from t^{\prime}\right)\tilde{p}_{jj^{\prime}}\left(t^{\prime}\right)\label{eq:CompositeSystemConditionalMarginalizationCondition}\\
 & \negmedspace\negmedspace\negmedspace\bracks{\textrm{for all }i\in\configspace,\,i^{\prime}\in\configspace^{\prime},\,t\in\tilde{\setoftimes},\,t^{\prime}\in\tilde{\setoftimes}_{0}}.\nonumber 
\end{align}
\end{itemize}

Given a composite system $\parens{\tilde{\configspace},\tilde{\setoftimes},\tilde{\setoftimes}_{0},\tilde{\stochasticmatrix},\tilde{p},\tilde{\rvalgebra}}$
with configuration space $\tilde{\configspace}=\configspace\cartesianprod\configspace^{\prime}$,
as in \eqref{eq:CompositeSystemConfigSpaceCartesianProd}, the composite
system's set of target times $\tilde{\setoftimes}$ trivially defines
two sets of target times $\setoftimes$ and $\setoftimes^{\prime}$
according to 
\begin{equation}
\setoftimes\defeq\setoftimes^{\prime}\defeq\tilde{\setoftimes},\label{eq:SubsystemSetsOfTargetTimes}
\end{equation}
 and the composite system's set of conditioning times $\tilde{\setoftimes}_{0}$
trivially defines two sets of conditioning times $\setoftimes_{0}$
and $\setoftimes_{0}^{\prime}$ according to 
\begin{equation}
\setoftimes_{0}\defeq\setoftimes_{0}^{\prime}\defeq\tilde{\setoftimes}_{0}.\label{eq:SubsystemSetsOfConditioningTimes}
\end{equation}
Meanwhile, the composite system's probability distribution $\tilde{p}$
defines two probability distributions $\mapping p{\configspace\cartesianprod\setoftimes}{\left[0,1\right]}$
and $\mapping{p^{\prime}}{\configspace^{\prime}\cartesianprod\setoftimes^{\prime}}{\left[0,1\right]}$
according to the respective marginalization formulas 
\begin{align}
p_{i}\left(t\right) & \defeq\sum_{i^{\prime}=1}^{N^{\prime}}\tilde{p}_{ii^{\prime}}\left(t\right)\quad\bracks{\textrm{for all }i\in\configspace,\,t\in\setoftimes},\label{eq:StandaloneProbabilityFirstSubsystemFromMaginalization}\\
p_{i^{\prime}}^{\prime}\left(t\right) & \defeq\sum_{i=1}^{N}\tilde{p}_{ii^{\prime}}\left(t\right)\quad\bracks{\textrm{for all }i^{\prime}\in\configspace^{\prime},\,t\in\setoftimes^{\prime}}.\label{eq:StandaloneProbabilitySecondSubsystemFromMaginalization}
\end{align}
 Note that these two probability distributions $p$ and $p^{\prime}$
do not necessarily turn $\configspace$ and $\configspace^{\prime}$
into indivisible stochastic processes of their own, in the sense of
\eqref{eq:DefIndivisibleStochasticSystem}, due to the generic lack
of well-defined transition maps $\mapping{\stochasticmatrix}{\configspace^{2}\cartesianprod\setoftimes\cartesianprod\setoftimes_{0}}{\left[0,1\right]}$
and $\mapping{\stochasticmatrix^{\prime}}{\configspace^{\prime2}\cartesianprod\setoftimes^{\prime}\cartesianprod\setoftimes_{0}^{\prime}}{\left[0,1\right]}$.
Indeed, in place of the law of total probability \eqref{eq:ConditionalBayesianMarginalizationConditionForGenericConditioningTimes},
one instead has the relations 
\begin{align}
p_{i}\left(t\right) & \defeq\sum_{i^{\prime}=1}^{N^{\prime}}\sum_{j=1}^{N}\sum_{j^{\prime}=1}^{N^{\prime}}\tilde{\stochasticmatrix}_{ii^{\prime},jj^{\prime}}\left(t\from t^{\prime}\right)\tilde{p}_{jj^{\prime}}\left(t^{\prime}\right)\label{eq:StandaloneProbabilityFirstSubsystemFromBayesianMarginalization}\\
 & \quad\qquad\bracks{\textrm{for all }i\in\configspace,\,t\in\setoftimes,\,t^{\prime}\in\setoftimes_{0}},\nonumber \\
p_{i^{\prime}}^{\prime}\left(t\right) & \defeq\sum_{i=1}^{N}\sum_{j=1}^{N}\sum_{j^{\prime}=1}^{N^{\prime}}\tilde{\stochasticmatrix}_{ii^{\prime},jj^{\prime}}\left(t\from t^{\prime}\right)\tilde{p}_{jj^{\prime}}\left(t^{\prime}\right)\label{eq:StandaloneProbabilitySecondSubsystemFromBayesianMarginalization}\\
 & \quad\qquad\bracks{\textrm{for all }i^{\prime}\in\configspace^{\prime},\,t\in\setoftimes^{\prime},\,t^{\prime}\in\setoftimes_{0}^{\prime}}.\nonumber 
\end{align}
 The two sets $\configspace$ and $\configspace^{\prime}$ will be
called the configuration spaces of subsystems of the composite system
$\parens{\tilde{\configspace},\tilde{\setoftimes},\tilde{\setoftimes}_{0},\tilde{\stochasticmatrix},\tilde{p},\tilde{\rvalgebra}}$.

\subsection{Unistochastic Processes\label{subsec:Unistochastic-Processes}}

Recall that an $N\times N$ matrix $U$ with complex-valued entries
is called unitary if it satisfies 
\begin{equation}
U^{\adj}U=UU^{\adj}=\idmatrix.\label{eq:DefUnitaryMatrix}
\end{equation}
 Here $\adj$ denotes the adjoint operation, meaning complex conjugation
combined with the transpose operation.

A stochastic matrix $X$ is called unistochastic if each of its entries
$X_{ij}$ is the modulus-square $\verts{U_{ij}}^{2}$ of the corresponding
entry $U_{ij}$ of a unitary matrix $U$: 
\begin{equation}
X=\begin{pmatrix}X_{11} & X_{12}\\
X_{21} & \ddots\\
 &  & X_{NN}
\end{pmatrix}=\begin{pmatrix}\verts{U_{11}}^{2} & \verts{U_{12}}^{2}\\
\verts{U_{21}}^{2} & \ddots\\
 &  & \verts{U_{NN}}^{2}
\end{pmatrix}.\label{eq:DefUnistochasticMatrix}
\end{equation}
 The study of unistochastic matrices was initiated in a 1954 paper
(Horn 1954)\nocite{Horn:1954dsmatdoarm}, which originally called
them \textquoteleft ortho-stochastic matrices.\textquoteright{} Unistochastic
matrices were given their modern name in 1989 (Thompson 1989; Nylen,
Tam, Hulig 1993; Bengtsson 2004)\nocite{Thompson:1989uln,NylenTamUhlig:1993oteopsonhasm,Bengtsson:2004tiobu},
and today orthostochastic matrices refer to the special case in which
$U$ can be taken to be a real-orthogonal matrix, meaning a unitary
matrix with entries that are all real-valued.

Every unistochastic matrix is doubly stochastic or bistochastic, meaning
that its columns\emph{ and} its rows each sum to $1$.\footnote{Proof: $\sum_{i}\verts{V_{ij}}^{2}=\sum_{i}\overconj{V_{ij}}V_{ij}=\bracks{V^{\adj}V}_{jj}=1$
and $\sum_{j}\verts{V_{ij}}^{2}=\sum_{j}V_{ij}\overconj{V_{ij}}=\bracks{VV^{\adj}}_{ii}=1$,
where overlines denote complex conjugation. QED} All $2\times2$ doubly stochastic matrices are unistochastic,\footnote{Proof: Every $2\times2$ doubly stochastic matrix is of the form $\left(\begin{smallmatrix}x & 1-x\\
1-x & x
\end{smallmatrix}\right)$, where $0\leq x\leq1$, and the matrix $\left(\begin{smallmatrix}\sqrt{x} & -\sqrt{1-x}\\
\sqrt{1-x} & \sqrt{x}
\end{smallmatrix}\right)$ is easily seen to be unitary. QED} but this equivalence does not extend beyond the $2\times2$ case.\footnote{For example, the $3\times3$ doubly stochastic matrix $\frac{1}{2}\left(\begin{smallmatrix}0 & 1 & 1\\
1 & 0 & 1\\
1 & 1 & 0
\end{smallmatrix}\right)$ is provably not unistochastic.} Moreover, beyond the $2\times2$ case, not every unistochastic matrix
is orthostochastic.\footnote{For example, the $3\times3$ matrix $\frac{1}{\sqrt{3}}\left(\begin{smallmatrix}1 & 1 & 1\\
1 & z & z^{2}\\
1 & z^{2} & z
\end{smallmatrix}\right)$ is unitary for $z=\exp\left(2\pi i/3\right)$, so the $3\times3$
matrix $\frac{1}{3}\left(\begin{smallmatrix}1 & 1 & 1\\
1 & 1 & 1\\
1 & 1 & 1
\end{smallmatrix}\right)$ is unistochastic. But this unistochastic matrix is not orthostochastic,
because there does not exist a set of three mutually orthogonal $3\times1$
column vectors that feature only $1$s and $-1$s as entries.}

An indivisible stochastic process $\left(\configspace,\setoftimes,\setoftimes_{0},\stochasticmatrix,p,\rvalgebra\right)$
whose transition map $\stochasticmatrix$ defines a unistochastic
matrix $\stochasticmatrix\left(t\from t_{0}\right)$ at every target
time $t$ and conditioning time $t_{0}$ will be called a unistochastic
process. One of the main goals of this paper will be to establish
that the study of indivisible stochastic processes essentially reduces
to the study of unistochastic processes, which will also turn out
to correspond to unitarily evolving quantum systems.

\section{The Stochastic-Quantum Theorem\label{sec:The-Stochastic-Quantum Theorem}}

\subsection{Statement of the Theorem\label{subsec:Statement-of-the-Theorem}}

Focusing on the case of indivisible stochastic processes with finite
configuration spaces (leaving the more general case to future work),
this paper's next major goal will be to present a self-contained proof
of the following theorem, which is a new result that implicitly also
appeared in other work (Barandes 2025)\nocite{Barandes:2025tsqc}.

\begin{equation}
\keyeq{\substack{{\displaystyle \textbf{The Stochastic-Quantum Theorem}}\\
\ \\
{\displaystyle \textrm{Every indivisible stochastic process}}\\
{\displaystyle \textrm{can be regarded as a subsystem of a}}\\
{\displaystyle \textrm{unistochastic process.}}
}
}\label{eq:StochasticQuantumTheorem}
\end{equation}
 Remarkably, to study the class of indivisible stochastic process,
this theorem implies that it suffices to restrict one's attention
to the subclass of unistochastic processes. 

\subsection{Corollaries of the Theorem\label{subsec:Corollaries-of-the-Theorem}}

The proof of the stochastic-quantum theorem \eqref{eq:StochasticQuantumTheorem}
will involve the construction of a representation of the given indivisible
stochastic process in the formalism of Hilbert spaces, and will show
that every indivisible stochastic process corresponds to a unitarily
evolving quantum system in a Hilbert space. One thereby turns some
of the puzzling axiomatic ingredients of quantum theory\textemdash the
complex numbers,\footnote{Time-reversal transformations for quantum systems are carried out
by anti-unitary operators, which take the general form $VK$, where
$V$ is unitary and $K$ is an abstract operator that carries out
complex-conjugation. By definition, $K^{2}=1$ and $Ki=-iK$, where
$i\defeq\sqrt{-1}$ is the imaginary unit. One can therefore show
that $i$, $K$, and $iK$ generate a Clifford algebra called the
pseudo-quaternions (Stueckelberg 1960)\nocite{Stueckelberg:1960qtirhs}.
As a consequence, there is a sense in which quantum systems are actually
defined over the pseudo-quaternions, rather than merely over the complex
numbers, although only the complex numbers are typically used in the
construction of observables.} Hilbert spaces, linear-unitary time evolution, and the Born rule
in particular\textemdash into the output of a theorem.

It follows from the stochastic-quantum theorem that all indivisible
stochastic processes (at least those based on finite configuration
spaces) can be modeled in terms of unitarily evolving quantum systems.
From this perspective, unitarily evolving quantum systems actually
represent the most general way to model a system with stochastic dynamical
laws.

As shown in other work (Barandes 2025)\nocite{Barandes:2025tsqc},
one can go in the other logical direction and show that any comprehensive
quantum system that includes measuring devices and observers as part
of the system can be modeled as an indivisible stochastic process.
Altogether, one thereby arrives at an important new correspondence
between indivisible stochastic processes and quantum systems, called
the stochastic-quantum correspondence, that provides a new way to
think about the physical meaning of quantum theory.

\subsection{A Simple Example\label{subsec:A-Simple-Example}}

The notion of embedding an indivisible stochastic process into a unistochastic
process does not always require an  elaborate construction. As a
simple example, consider a discrete Markovian-homogeneous dynamical
system $\left(\statespace,\setoftimes,f\right)$ whose state space
$\statespace$ has some finite size $N$, whose set of target times
$\setoftimes$ is isomorphic to the integers $\mathbb{Z}$ under addition,
and whose dynamical map $\mapping f{\statespace\cartesianprod\setoftimes}{\statespace}$
satisfies the appropriate version of the homogeneity property \eqref{eq:DynamicalSystemSemigroupActionHomogeneityProperty}:
\begin{equation}
f_{t+t^{\prime}}=f_{t}\composition f_{t^{\prime}}\quad\left[\textrm{for all }t,t^{\prime}\in\setoftimes\cong\mathbb{Z}\right].\label{eq:DiscreteTimeDynamicalSystemSemigroupActionMarkovProperty}
\end{equation}
 Based on these definitions and assumptions, it follows from the
homogeneity property and the trivialization condition $f_{0}=\mathrm{id}_{\statespace}$
from \eqref{eq:IndivisibleDynamicalSystemDynamicalMapInitialTimeTrivial}
that the time-dependent dynamical map $f_{t}$ is invertible for any
time $t\in\setoftimes$, with inverse given by 
\begin{equation}
f_{t}^{-1}=f_{-t}\quad\left[\textrm{for all }t\in\setoftimes\cong\mathbb{Z}\right].\label{eq:DiscreteTimeDynamicalSystemInverseTimeDepDynamicalMap}
\end{equation}
 Markovian-homogeneous dynamical systems of this kind provide a discretization
of many of the kinds of deterministic, logically time-reversible systems
that show up in classical physics.

Letting $\delta t$ be the elementary duration of a single time step,
and letting $n\in\mathbb{Z}$ denote an integer number of time steps,
the system's specific state $i$ at the time $t=n\,\delta t$ can
be identified with an $N\times1$ column vector with a $1$ in its
$i$th entry and $0$s in all its other entries. Moreover, the time-dependent
dynamical map $f_{n\,\delta t}$ can be expressed as the $n$th power
of a fixed permutation matrix $\permutationmatrix$, meaning a matrix
consisting of only $1$s and $0$s, with a single $1$ in each row
and in each column.

Every permutation matrix is unitary, and is also unistochastic, because
computing the modulus-squares of $1$s and $0$s gives back $1$s
and $0$s. Hence, a discrete Markovian-homogeneous dynamical system
defined in this way is already a unistochastic process, and therefore
trivially satisfies the stochastic-quantum theorem \eqref{eq:StochasticQuantumTheorem}.

Interestingly, there exists an analytic interpolation of this \emph{discrete-time}
unistochastic process that yields a corresponding \emph{continuous-time}
unistochastic process, at the cost of introducing nontrivial stochasticity
into the intervals between the discrete time steps. This new unistochastic
process is based on a time-dependent unistochastic matrix $\stochasticmatrix\left(t\from0\right)$
whose entries are the modulus-squares of the corresponding entries
of the $N\times N$ unitary time-evolution operator defined by $\timeevop\left(t\from0\right)\defeq\permutationmatrix^{t/\delta t}$,
and therefore satisfies $\stochasticmatrix\left(n\,\delta t\from0\right)=\permutationmatrix^{n}$
for any integer $n$. One can even go on to define an $N\times N$
self-adjoint Hamiltonian $H$ as the infinitesimal generator of $\timeevop\left(t\from0\right)$,
so there is ultimately a Schrödinger equation underlying this system.\footnote{To define the unitary matrix $\permutationmatrix^{t/\delta t}$, one
starts by using the fact that every permutation matrix is also unitary
to write the permutation matrix $\permutationmatrix$ as $V^{\adj}DV$
for some unitary matrix $V$ and some diagonal matrix $D$, where
the entries of $D$ are the eigenvalues of $\permutationmatrix$.
Because $\permutationmatrix$ is unitary, its eigenvalues are all
phase factors $\exp\left(i\theta_{m}\right)$, where $m=1,\dots,N$,
and where each phase $\theta_{m}$ is a real number. Setting all the
eigenvalues to the power $t/\delta t$, so that they each take the
new form $\exp\left(i\theta_{m}t/\delta t\right)$, one ends up with
the matrix $\permutationmatrix^{t/\delta t}\defeq V^{\adj}D^{t/\delta t}V$,
which is still unitary and now depends analytically on the time parameter
$t$. The quantum-theoretic Hamiltonian matrix $H$ for this system
then has energy eigenvalues defined by $E_{m}\defeq-\hbar\theta_{m}/\delta t$
for $m=1,\dots,N$, and is diagonalized by the same unitary matrix
$V$ that diagonalizes $\permutationmatrix$.} 

\section{Proof of the Theorem\label{sec:Proof-of-the-Theorem}}

\subsection{The Time-Evolution Operator\label{subsec:The-Time-Evolution-Operator}}

To commence the proof of the stochastic-quantum theorem \eqref{eq:StochasticQuantumTheorem},
one starts with a given indivisible stochastic process $\left(\configspace,\setoftimes,\setoftimes_{0},\stochasticmatrix,p,\rvalgebra\right)$
with a finite configuration space $\configspace$ of size $N$. Singling
out one conditioning time $t_{0}$, which will be taken to be the
initial time $0$ for convenience, the non-negativity \eqref{eq:TransitionProbabilitiesNonNegative}
of the system's conditional transition probabilities, $\stochasticmatrix_{ij}\left(t\from0\right)\geq0,$
means that each transition probability can be written as the modulus-square
of a non-unique complex number $\dynop_{ij}\left(t\from0\right)$:
\begin{equation}
\stochasticmatrix_{ij}\left(t\from0\right)=\absval{\dynop_{ij}\left(t\from0\right)}^{2}\quad\left[\textrm{for all }i,j\in\configspace,\,t\in\setoftimes\right].\label{eq:StochasticMatrixEntryFromAbsValSqTimeEvOp}
\end{equation}
 It is worth emphasizing that this formula is an identity, not a postulate.
Any non-negative real number can be written non-uniquely as the modulus-square
of a complex number.

For each fixed target time $t$, the complex numbers $\dynop_{ij}\left(t\from0\right)$
collectively form their own $N\times N$ matrix $\dynop\left(t\from0\right)$,
which one can think of as a non-unique \textquoteleft potential matrix\textquoteright{}
for $\stochasticmatrix_{ij}\left(t\from0\right)$, and will be called
the system's time-evolution operator: 
\begin{equation}
\dynop\left(t\from0\right)\defeq\begin{pmatrix}\dynop_{11}\left(t\from0\right) & \dynop_{12}\left(t\from0\right)\\
\dynop_{21}\left(t\from0\right) & \ddots\\
 &  & \dynop_{NN}\left(t\from0\right)
\end{pmatrix}.\label{eq:TimeEvOpAsMatrix}
\end{equation}
 (As a concession to terminological conventions, the terms \textquoteleft matrix\textquoteright{}
and \textquoteleft operator\textquoteright{} will be used more-or-less
interchangeably in what follows.)

The normalization condition \eqref{eq:ConditionalProbabilityNormalization}
on the system's transition matrix $\stochasticmatrix\left(t\from0\right)$
then becomes the summation condition 
\begin{equation}
\sum_{i=1}^{N}\absval{\dynop_{ij}\left(t\from0\right)}^{2}=1\quad\left[\textrm{for all }j\in\configspace,\,t\in\setoftimes\right],\label{eq:TimeEvOpSummationCondition}
\end{equation}
 which can roughly be regarded as a generalization of a unitarity
constraint.  In keeping with the trivialization condition $\stochasticmatrix\left(0\from0\right)=\idmatrix$
from \eqref{eq:InitialConditionStochasticMatrix}, the time-evolution
operator $\dynop\left(0\from0\right)$ for $t=0$ will be taken to
be the $N\times N$ identity matrix $\idmatrix$: 
\begin{equation}
\dynop\left(0\from0\right)\defeq\idmatrix\defeq\begin{pmatrix}1 & 0\\
0 & \ddots\\
 &  & 1
\end{pmatrix}.\label{eq:TimeEvOpInitialCondition}
\end{equation}

\subsection{The Hilbert Space\label{subsec:The-Hilbert-Space}}

The time-evolution operator $\dynop\left(t\from0\right)$ acts on
an $N$-dimensional Hilbert space $\hilbspace$ defined as the space
$\mathbb{C}^{N}$ of $N\times1$ column vectors with complex-valued
entries, 
\begin{equation}
\hilbspace\defeq\mathbb{C}^{N},\label{eq:DefHilbertSpace}
\end{equation}
 together with the standard inner product $v^{\adj}w$ for all $v,w\in\mathbb{C}^{N}$.
The standard orthonormal basis $e_{1},\dots,e_{N}$ is defined by
\begin{equation}
e_{1}\defeq\left(\begin{smallmatrix}1\\
0\\
\vdots\\
0\\
0
\end{smallmatrix}\right),\quad\dots,\quad e_{N}\defeq\left(\begin{smallmatrix}0\\
0\\
\vdots\\
0\\
1
\end{smallmatrix}\right).\label{eq:DefConfigurationBasis}
\end{equation}
 These vectors are labeled as $e_{i}$, where each value of $i=1,\dots,N$
denotes a configuration in the system's configuration space $\configspace$,
so this basis will be called the system's configuration basis. 

There exists an associated set of rank-one projectors $\projector_{1},\dots,\projector_{N}$
defined by 
\begin{equation}
\projector_{i}\defeq e_{i}e_{i}^{\adj}=\mathrm{diag}\parens{0,\dots,0,\underset{\mathclap{\substack{\uparrow\\
i\textrm{th entry}
}
}}{1},0,\dots,0}\quad\left[\textrm{for all }i\in\configspace\right],\label{eq:DefConfigurationPVM}
\end{equation}
 and called configuration projectors. These configuration projectors
$\projector_{1},\dots,\projector_{N}$ form a projection-valued measure
(PVM) (Mackey 1952, 1957)\nocite{Mackey:1952irolcgi,Mackey:1957qmahs},
meaning that they satisfy the conditions of mutual exclusivity, 
\begin{equation}
\projector_{i}\projector_{j}=\delta_{ij}\projector_{i}\quad\left[\textrm{for all }i,j\in\configspace\right],\label{eq:ConfigurationPVMMutualExclusivity}
\end{equation}
 with $\delta_{ij}$ again the usual Kronecker delta, and completeness,
\begin{equation}
\sum_{i=1}^{N}\projector_{i}=\idmatrix,\label{eq:ConfigurationPVMCompleteness}
\end{equation}
 with $\idmatrix$ again the $N\times N$ identity matrix.

\subsection{The Dictionary\label{subsec:The-Dictionary}}

It follows from a short calculation that the relationship $\stochasticmatrix_{ij}\left(t\from0\right)=\verts{\dynop_{ij}\left(t\from0\right)}^{2}$
in the identity \eqref{eq:StochasticMatrixEntryFromAbsValSqTimeEvOp}
can be expressed in terms of a matrix trace as 
\begin{align}
 & \keyeq{\stochasticmatrix_{ij}\left(t\from0\right)=\tr\bigparens{\dynop^{\adj}\left(t\from0\right)\projector_{i}\dynop\left(t\from0\right)\projector_{j}}}\label{eq:StochasticQuantumDictionary}\\
 & \quad\qquad\qquad\left[\textrm{for all }i,j\in\configspace,\,t\in\setoftimes\right].\nonumber 
\end{align}
 This \textquoteleft dictionary\textquoteright{} essentially translates
between the theory of indivisible stochastic processes, as expressed
by the left-hand side, and a corresponding Hilbert-space representation,
as expressed by the right-hand side. This equation turns out to provide
the foundation for a stochastic-quantum correspondence (Barandes
2025)\nocite{Barandes:2025tsqc} and will play a crucial role in the
proof of the stochastic-quantum theorem \eqref{eq:StochasticQuantumTheorem}.

\subsection{The Density Matrix\label{subsec:The-Density-Matrix}}

Fixing the conditioning time $t_{0}$ to be the initial time $0$
and inserting the dictionary \eqref{eq:StochasticQuantumDictionary}
into the law of total probability \eqref{eq:ConditionalBayesianMarginalizationCondition}
yields the following equation: 
\begin{equation}
p_{i}\left(t\right)=\tr\left(\projector_{i}\densitymatrix\left(t\right)\right)\quad\left[\textrm{for all }i\in\configspace,\,t\in\setoftimes\right].\label{eq:FinalStandaloneProbabilityFromTraceProjectionDensityMatrix}
\end{equation}
 Here $\densitymatrix\left(t\right)$, which will be called the system's
density matrix, is an $N\times N$ time-dependent, self-adjoint, unit-trace,
generically non-diagonal matrix defined for any target time $t$ by
\begin{equation}
\densitymatrix\left(t\right)\defeq\dynop\left(t\from0\right)\densitymatrix\left(0\right)\dynop^{\adj}\left(t\from0\right),\label{eq:DefDensityMatrix}
\end{equation}
 where its value at the initial time $0$ is the following $N\times N$
diagonal matrix: 
\begin{equation}
\densitymatrix\left(0\right)\defeq\sum_{j=1}^{N}p_{j}\left(0\right)\projector_{j}=\diag{p_{1}\left(0\right),\dots,p_{N}\left(0\right)}=\begin{pmatrix}p_{1}\left(0\right) & 0\\
0 & \ddots\\
 &  & p_{N}\left(0\right)
\end{pmatrix}.\label{eq:DefInitialDensityMatrix}
\end{equation}
 Importantly, one sees from this analysis that the linearity of the
law of total probability \eqref{eq:ConditionalBayesianMarginalizationCondition}
underlies the linearity of the relationship \eqref{eq:DefDensityMatrix}
between the system's density matrix $\densitymatrix\left(0\right)$
at the initial time $0$ and its density matrix $\densitymatrix\left(t\right)$
at target times $t$.

For any target time $t$, one can similarly express the expectation
value \eqref{eq:DefExpectationValueRandomVariable} of a random variable
$A$ as 
\begin{equation}
\expectval{A\left(t\right)}=\tr\left(A\left(t\right)\densitymatrix\left(t\right)\right).\label{eq:ExpectationValueAsTrace}
\end{equation}
 Here $A\left(t\right)$ denotes the $N\times N$ diagonal matrix
\begin{equation}
A\left(t\right)\defeq\sum_{i=1}^{N}a_{i}\left(t\right)\projector_{i}=\diag{a_{1}\left(t\right),\dots,a_{N}\left(t\right)}=\begin{pmatrix}a_{1}\left(t\right) & 0\\
0 & \ddots\\
 &  & a_{N}\left(t\right)
\end{pmatrix}.\label{eq:DefRandomVariableAsMatrix}
\end{equation}
 Observe, in particular, that the magnitudes $a_{1}\left(t\right),\dots,a_{N}\left(t\right)$
of the random variable $A$ become the eigenvalues of the $N\times N$
matrix $A\left(t\right)$.

Notice also that the formula for the standalone probability $p_{i}\left(t\right)$
in \eqref{eq:FinalStandaloneProbabilityFromTraceProjectionDensityMatrix}
and the formula for the expectation value $\expectval{A\left(t\right)}$
in \eqref{eq:ExpectationValueAsTrace} are both special cases of the
Born rule.

\subsection{An Aside on \textquoteleft Classical Wave Functions\textquoteright \label{subsec:An-Aside-on-Classical-Wave-Functions}}

Pausing for a moment, recall the smooth unistochastic interpolation
of a discrete Markovian-homogeneous dynamical system described in
Subsection~\ref{subsec:A-Simple-Example}. For that system, the unitary
time-evolution operator $\timeevop\left(t\from0\right)\defeq\permutationmatrix^{t/\delta t}$
trivializes to a permutation matrix $\permutationmatrix^{n}$ at every
integer time step $n\,\delta t$. It follows that the system's density
matrix $\densitymatrix\left(t\right)$, as defined in terms of its
initial value $\densitymatrix\left(0\right)$ from \eqref{eq:DefDensityMatrix},
reduces to a diagonal matrix $\densitymatrix\left(n\,\delta t\right)$
at each integer time step. Taking the square root of each of its diagonal
entries $p_{1}\left(n\,\delta t\right),\dots,p_{N}\left(n\,\delta t\right)$,
and allowing for arbitrary phase factors, one obtains a \textquoteleft classical
wave function\textquoteright{} with components $\Psi_{1}\left(n\,\delta t\right),\dots,\Psi_{N}\left(n\,\delta t\right)$
satisfying $\verts{\Psi_{i}\left(n\,\delta t\right)}^{2}=p_{i}\left(n\,\delta t\right)$
and capturing precisely the same information as the diagonal density
matrix $\densitymatrix\left(n\,\delta t\right)$.

This classical wave function is the starting place for a representation
of classical deterministic physics known popularly as the \textquoteleft Koopman-von
Neumann\textquoteright{} formulation, due to its superficial resemblance
to work by Koopman (1931)\nocite{Koopman:1931hsatihs} and von Neumann
(1932a, 1932b)\nocite{vonNeumann:1932zoidkm,vonNeumann:1932zzazo}
in the 1930s. However, as pointed out explicitly by Jordan and Sudarshan
(1961)\nocite{JordanSudarshan:1961dmdoqm}, and in other work (Barandes
2026)\nocite{Barandes:2026thohsfocp}, Koopman and von Neumann were
actually using Hilbert spaces to represent \emph{observables}, rather
than to represent \emph{probability distributions}. The formulation
of classical physics in terms of \textquoteleft classical wave functions\textquoteright{}
is more properly due to Schönberg (1953)\nocite{Schoenberg:1953aosqmttcsmii},
Loinger (1962)\nocite{Loinger:1962ggale}, Della Riccia and Wiener
(1966)\nocite{DellaRicciaWiener:1966wmicpsbmaqt}, and Sudarshan (1976)\nocite{Sudarshan:1976ibcaqsatmoqo}.

\subsection{The Kraus Decomposition\label{subsec:The-Kraus-Decomposition}}

Returning to the proof, fixing the target time $t$, and letting $\beta$
denote an integer from $1$ to $N$ that will play a conceptually
different role from a configuration index, let $\krausmatrix_{\beta}\left(t\from0\right)$
denote the $N\times N$ matrix whose $\beta$th column agrees with
the $\beta$th column of the time-evolution operator $\dynop\left(t\from0\right)$,
with $0$s in all its other entries: 
\begin{equation}
\krausmatrix_{\beta}\left(t\from0\right)=\begin{pmatrix}\begin{matrix}0 & \cdots & 0 & \dynop_{1\beta}\left(t\from0\right) & 0 & \cdots & 0\\
0 & \cdots & 0 & \dynop_{2\beta}\left(t\from0\right) & 0 & \cdots & 0\\
\vdots &  & \vdots & \vdots & \vdots &  & \vdots\\
0 & \cdots & 0 & \dynop_{N\beta}\left(t\from0\right) & 0 & \cdots & 0
\end{matrix}\end{pmatrix}.\label{eq:DefKrausOperators}
\end{equation}
 That is, the entry in the $i$th row, $j$th column of $\krausmatrix_{\beta}\left(t\from0\right)$
is given by 
\begin{equation}
\krausmatrix_{\beta,ij}\left(t\right)\defeq\left(\dynop\left(t\from0\right)\projector_{\beta}\right){}_{ij}\defeq\delta_{\beta j}\dynop_{ij}\left(t\right)\quad\left[\textrm{for all }\beta,i,j\in\configspace,\,t\in\setoftimes\right].\label{eq:DefKrausOperatorsEntry}
\end{equation}
 It follows from the summation condition \eqref{eq:TimeEvOpSummationCondition}
on $\dynop\left(t\from0\right)$ that these new matrices satisfy the
Kraus condition\index{Kraus condition}: 
\begin{equation}
\sum_{\beta=1}^{N}\krausmatrix_{\beta}^{\adj}\left(t\from0\right)\krausmatrix_{\beta}\left(t\from0\right)=\idmatrix\quad\left[\textrm{for all }t\in\setoftimes\right].\label{eq:KrausCondition}
\end{equation}
 Moreover, one can write the dictionary \eqref{eq:StochasticQuantumDictionary}
as 
\begin{align}
\stochasticmatrix_{ij}\left(t\from0\right) & =\sum_{\beta=1}^{N}\tr\bigparens{\krausmatrix_{\beta}^{\adj}\left(t\from0\right)\projector_{i}\krausmatrix_{\beta}\left(t\from0\right)\projector_{j}}\label{eq:DictionaryFromKrausOps}\\
 & \negthickspace\negthickspace\negthickspace\left[\textrm{for all }i,j\in\configspace,\,t\in\setoftimes\right],\nonumber 
\end{align}
 and one can express the time-evolution rule \eqref{eq:DefDensityMatrix}
for the system's density matrix as 
\begin{equation}
\densitymatrix\left(t\right)=\sum_{\beta=1}^{N}\krausmatrix_{\beta}\left(t\from0\right)\densitymatrix\left(0\right)\krausmatrix_{\beta}^{\adj}\left(t\from0\right).\label{eq:FinalDensityMatrixFromKrausDecomposition}
\end{equation}
 The matrices $\krausmatrix_{1}\left(t\from0\right),\dots,\krausmatrix_{N}\left(t\from0\right)$
are therefore Kraus operators (Kraus 1971)\nocite{Kraus:1971gscqt},
and \eqref{eq:FinalDensityMatrixFromKrausDecomposition} gives a Kraus
decomposition of $\densitymatrix\left(t\right)$.

Abstracting these results, one obtains a completely positive trace-preserving
(CPTP) map, or quantum channel, 
\begin{equation}
\mapping{\cptpmap[t\from0]}{\mathbb{C}^{N\times N}}{\mathbb{C}^{N\times N}},\label{eq:DefQuantumChannelLinearCPTPMap}
\end{equation}
 defined on the algebra $\mathbb{C}^{N\times N}$ of $N\times N$
matrices over the complex numbers according to 
\begin{equation}
X\mapsto\cptpmap[t\from0]\left(X\right)\defeq\sum_{\beta=1}^{N}\krausmatrix_{\beta}\left(t\from0\right)X\krausmatrix_{\beta}^{\adj}\left(t\from0\right).\label{eq:DefQuantumChannel}
\end{equation}

\subsection{Dilation\label{subsec:Dilation}}

By the Stinespring dilation theorem (Stinespring 1955, Keyl 2002)\nocite{Stinespring:1955pfoc,Keyl:2002foqit},
the quantum channel \eqref{eq:DefQuantumChannelLinearCPTPMap} can
be purified, meaning made into a form of unitary time evolution. Specifically,
for some integer $\tilde{N}$ in the \emph{bounded} interval 
\begin{equation}
N\leq\tilde{N}\leq N^{3},\label{eq:DilatedHilbertSpaceDimensionInterval}
\end{equation}
  where $\tilde{N}$ is an integer multiple of $N$, there exists
an $\tilde{N}\times\tilde{N}$ unitary matrix $\tilde{\timeevop}\left(t\from0\right)$
on a potentially enlarged or dilated Hilbert space 
\begin{equation}
\tilde{\hilbspace}\defeq\mathbb{C}^{\tilde{N}}\label{eq:DefDilatedHilbertSpace}
\end{equation}
 such that the dictionary \eqref{eq:StochasticQuantumDictionary}
can be written as\footnote{From the starting assumptions presented here, one can sketch the following
proof: Given $N\times N$ Kraus matrices $\krausmatrix_{\beta}\left(t\from0\right)$,
with $\beta=1,\dots,N$, define an $N^{3}\times N^{2}$ matrix $\tilde{V}\left(t\right)$
according to $\tilde{V}_{\left(i\beta m\right)\left(jl\right)}\left(t\from0\right)\defeq\krausmatrix_{\beta,ij}\left(t\from0\right)\delta_{lm}$,
treating $\left(i\beta m\right)$ as the first index of $\tilde{V}\left(t\from0\right)$
and treating $\left(jl\right)$ as its second index. One can show
that this matrix satisfies $\tilde{V}^{\adj}\left(t\from0\right)\tilde{V}\left(t\from0\right)=\idmatrix_{N^{2}\times N^{2}}$,
so it defines a partial isometry, which can always be extended to
a unitary $N^{3}\times N^{3}$ matrix $\tilde{\timeevop}_{\left(i\beta m\right)\left(ja\right)}\left(t\from0\right)$
by adding $N^{3}-N^{2}$ additional columns that are mutually orthogonal
with each other and with the previous $N^{2}$ columns already in
$\tilde{V}\left(t\from0\right)$, where the new index $a$ runs through
$N^{2}$ possible values. These additional columns can always be chosen
so that at the initial time $0$, where $V\left(0\from0\right)=\idmatrix$
is the $N\times N$ identity matrix, they make $\tilde{\timeevop}\left(0\from0\right)$
coincide with the $N^{3}\times N^{3}$ identity matrix. The last step
is to show that $\tilde{\timeevop}\left(t\from0\right)$ satisfies
\eqref{eq:DictionaryAfterDilation}, whose right-hand side reduces
to $\sum_{\beta,m}\verts{\tilde{\timeevop}_{\left(i\beta m\right)\left(jj^{\prime}\right)}\left(t\from0\right)}^{2}=\sum_{\beta}\verts{\krausmatrix_{\beta,ij}\left(t\from0\right)}^{2}$.~QED} 
\begin{align}
\stochasticmatrix_{ij}\left(t\from0\right) & =\tr\left(\tr^{\prime}\left(\tilde{\timeevop}^{\adj}\left(t\from0\right)\bracks{\projector_{i}\tensorprod\idmatrix^{\prime}}\tilde{\timeevop}\left(t\from0\right)\bracks{\projector_{j}\tensorprod\projector_{j^{\prime}}^{\prime}}\right)\right)\label{eq:DictionaryAfterDilation}\\
 & \quad\qquad\bracks{\textrm{for all }i,j\in\configspace,\,t\in\setoftimes}.\nonumber 
\end{align}

The formula \eqref{eq:DictionaryAfterDilation} involves a number
of ingredients.
\begin{itemize}
\item The dilated Hilbert space $\tilde{\hilbspace}$ is defined as the
tensor product 
\begin{equation}
\tilde{\hilbspace}\defeq\hilbspace\tensorprod\hilbspace^{\prime},\label{eq:DilatedHilbertSpaceTensorFactorized}
\end{equation}
 where $\hilbspace\defeq\mathbb{C}^{N}$ is the system's original
Hilbert space \eqref{eq:DefHilbertSpace}, and where 
\begin{equation}
\hilbspace^{\prime}\defeq\mathbb{C}^{N^{\prime}}\label{eq:DefAncillaryHilbertSpace}
\end{equation}
 is an ancillary Hilbert space whose dimension $N^{\prime}$ satisfies
$\tilde{N}=NN^{\prime}\leq N^{3}$. That is, $N^{\prime}$ is an integer
lying in the bounded interval 
\begin{equation}
1\leq N^{\prime}\leq N^{2}.\label{eq:AncillaryHilbertSpaceDimensionInterval}
\end{equation}
\item The first trace $\tr\left(\cdots\right)$ in \eqref{eq:DictionaryAfterDilation}
denotes the partial trace over just the original Hilbert space $\hilbspace$,
and the second trace $\tr^{\prime}\left(\cdots\right)$ similarly
denotes the partial trace over the ancillary Hilbert space $\hilbspace^{\prime}$.
\item Importantly, for each fixed time $t$, $\tilde{\timeevop}\left(t\from0\right)$
is an $\tilde{N}\times\tilde{N}$ unitary matrix that reduces to the
$\tilde{N}\times\tilde{N}$ identity matrix $\tilde{\timeevop}\left(0\from0\right)=\tilde{\idmatrix}$
at the initial time $0$.
\item The symbol $\idmatrix^{\prime}$ denotes the $N^{\prime}\times N^{\prime}$
identity matrix on the ancillary Hilbert space $\hilbspace^{\prime}$.
\item Letting $e_{1}^{\prime},\dots,e_{N^{\prime}}^{\prime}$ denote the
standard orthonormal basis for the ancillary Hilbert space $\hilbspace^{\prime}$,
in analogy with the configuration basis \eqref{eq:DefConfigurationBasis}
for the system's original Hilbert space $\hilbspace$, and letting
the primed Latin letters $i^{\prime},j^{\prime},\dots$ each denote
an element of an ancillary configuration space $\configspace^{\prime}$
consisting of the integers from $1$ to $N^{\prime}$, the symbol
$\projector_{i^{\prime}}^{\prime}$ denotes the rank-one projector
\begin{align}
\projector_{i^{\prime}}^{\prime} & \defeq e_{i^{\prime}}^{\prime}e_{i^{\prime}}^{\prime\adj}=\mathrm{diag}\parens{0,\dots,0,\underset{\mathclap{\substack{\uparrow\\
i^{\prime}\textrm{th entry}
}
}}{1},0,\dots,0}\label{eq:DefAncillaryConfigurationPVM}\\
 & \quad\left[\textrm{for all }i^{\prime}\in\configspace^{\prime}\right],\nonumber 
\end{align}
 which is an $N^{\prime}\times N^{\prime}$ diagonal matrix with a
$1$ in its $i^{\prime}$th diagonal entry and $0$s in all its other
entries. These projectors form a projection-valued measure (PVM) on
$\hilbspace^{\prime}$ satisfying the conditions of mutual exclusivity,
\begin{equation}
\projector_{i^{\prime}}^{\prime}\projector_{j^{\prime}}^{\prime}=\delta_{i^{\prime}j^{\prime}}\projector_{i^{\prime}}^{\prime}\quad\left[\textrm{for all }i^{\prime},j^{\prime}\in\configspace^{\prime}\right],\label{eq:AncillaryConfigurationPVMMutualExclusivity}
\end{equation}
 and completeness, 
\begin{equation}
\sum_{i^{\prime}=1}^{N^{\prime}}\projector_{i^{\prime}}^{\prime}=\idmatrix^{\prime}.\label{eq:AncillaryConfigurationPVMCompleteness}
\end{equation}
\item Note that the left-hand side of \eqref{eq:DictionaryAfterDilation}
is insensitive to the specific choice of $j^{\prime}\in\configspace^{\prime}$
on the right-hand side. 
\end{itemize}

Extending the foregoing construction, and fixing the target time $t$,
one obtains an $\tilde{N}\times\tilde{N}$ transition matrix given
by the new dictionary 
\begin{align}
\tilde{\stochasticmatrix}_{ii^{\prime},jj^{\prime}}\left(t\from0\right) & \defeq\tilde{\tr}\bigparens{\tilde{\timeevop}^{\adj}\left(t\from0\right)\tilde{\projector}_{ii^{\prime}}\tilde{\timeevop}\left(t\from0\right)\tilde{\projector}_{jj^{\prime}}}\label{eq:DefDilatedUnistochasticMatrix}\\
 & \negthickspace\negthickspace\negthickspace\left[\textrm{for all }i,j\in\configspace,\,i^{\prime},j^{\prime}\in\configspace^{\prime},\,t\in\setoftimes\right].\nonumber 
\end{align}
 Here the trace is now over the dilated Hilbert space $\tilde{\hilbspace}$
defined in \eqref{eq:DefDilatedHilbertSpace}, and 
\begin{equation}
\tilde{\projector}_{ii^{\prime}}\defeq\projector_{i}\tensorprod\tilde{\projector}_{i^{\prime}}\quad\left[\textrm{for all }i\in\configspace,\,i^{\prime}\in\configspace^{\prime}\right]\label{eq:DefDilatedConfigurationPVM}
\end{equation}
 defines a rank-one projector on $\tilde{\hilbspace}$.

Unfolding the notation, the dilated dictionary \eqref{eq:DefDilatedUnistochasticMatrix}
reduces to the statement that 
\begin{align}
\tilde{\stochasticmatrix}_{ii^{\prime},jj^{\prime}}\left(t\from0\right) & =\verts{\tilde{\timeevop}_{ii^{\prime},jj^{\prime}}\left(t\from0\right)}^{2}\label{eq:DilatedUnistochasticMatrixFromAbsValSq}\\
 & \negthickspace\negthickspace\negthickspace\left[\textrm{for all }i,j\in\configspace,\,i^{\prime},j^{\prime}\in\configspace^{\prime},\,t\in\setoftimes\right].\nonumber 
\end{align}
 That is, each entry $\tilde{\stochasticmatrix}_{ii^{\prime},jj^{\prime}}\left(t\from0\right)$
of the $\tilde{N}\times\tilde{N}$ transition matrix $\tilde{\stochasticmatrix}\left(t\right)$
is the modulus-square of the corresponding entry $\tilde{\timeevop}_{ii^{\prime},jj^{\prime}}\left(t\from0\right)$
of an $\tilde{N}\times\tilde{N}$ unitary matrix $\tilde{\timeevop}\left(t\from0\right)$.
Again, a stochastic matrix with this special feature is called a unistochastic
matrix.

\subsection{The Dilated Indivisible Stochastic Process\label{subsec:The-Dilated-Indivisible-Stochastic-Process}}

One can now define a dilated indivisible stochastic process $\parens{\tilde{\configspace},\tilde{\setoftimes},\tilde{\setoftimes}_{0},\tilde{\stochasticmatrix},\tilde{p},\tilde{\rvalgebra}}$
in the following way.
\begin{itemize}
\item Let the dilated system's configuration space $\tilde{\configspace}$
be defined as the Cartesian product 
\begin{equation}
\tilde{\configspace}\defeq\configspace\cartesianprod\configspace^{\prime},\label{eq:DefDilatedConfigSpace}
\end{equation}
 where $\configspace$ is the original system's configuration space,
with $N$ elements labeled by unprimed Latin letters $i,j,\dots$,
and where $\configspace^{\prime}$ is the configuration space of an
ancillary subsystem, with $N^{\prime}$ elements labeled by primed
Latin letters $i^{\prime},j^{\prime},\dots$.
\item Let the dilated system's set of target times $\tilde{\setoftimes}$
be the original system's set of target times $\setoftimes$: 
\begin{equation}
\tilde{\setoftimes}\defeq\setoftimes.\label{eq:DilatedSetOfTimes}
\end{equation}
\item Let the dilated system's set of conditioning times $\tilde{\setoftimes}_{0}$
be the singleton set $\left\{ 0\right\} $.
\item Let $\mapping{\tilde{\stochasticmatrix}}{\tilde{\configspace}^{2}\cartesianprod\tilde{\setoftimes}\cartesianprod\left\{ 0\right\} }{\left[0,1\right]}$
be the transition map defined according to the dilated dictionary
\eqref{eq:DilatedUnistochasticMatrixFromAbsValSq}. Then for each
target time $t$, the $\tilde{N}\times\tilde{N}$ matrix $\tilde{\stochasticmatrix}\left(t\from0\right)$
is unistochastic. Moreover, by construction, $\tilde{\stochasticmatrix}\left(t\from0\right)$
satisfies the marginalization condition 
\begin{align}
\sum_{i^{\prime}=1}^{N^{\prime}}\tilde{\stochasticmatrix}_{ii^{\prime},jj^{\prime}}\left(t\from0\right) & =\stochasticmatrix_{ij}\left(t\from0\right)\label{eq:DilatedUnistochasticMatrixMarginalizationCondition}\\
 & \negthickspace\negthickspace\negthickspace\left[\textrm{for all }i,j\in\configspace,\,j^{\prime}\in\configspace^{\prime},\,t\in\setoftimes\right],\nonumber 
\end{align}
 where, as in \eqref{eq:DictionaryAfterDilation}, the value of $j^{\prime}\in\configspace^{\prime}$
is irrelevant. 
\item Let the map $\mapping{\tilde{p}}{\tilde{\configspace}}{\left[0,1\right]}$
be the probability distribution defined by 
\begin{align}
\tilde{p}_{ii^{\prime}}\left(t\right) & \defeq\sum_{j=1}^{N}\tilde{\stochasticmatrix}_{ii^{\prime},jj^{\prime}}\left(t\from0\right)p_{j}\left(0\right)\label{eq:DefDilatedProbabilityDistribution}\\
 & \negthickspace\negthickspace\negthickspace\left[\textrm{for all }i\in\configspace,\,i^{\prime},j^{\prime}\in\configspace^{\prime},\,t\in\setoftimes\right].\nonumber 
\end{align}
It follows from the marginalization condition \eqref{eq:DilatedUnistochasticMatrixMarginalizationCondition},
together with the law of total probability \eqref{eq:ConditionalBayesianMarginalizationCondition},
that 
\begin{equation}
\sum_{i^{\prime}=1}^{N^{\prime}}\tilde{p}_{ii^{\prime}}\left(t\right)=p_{i}\left(t\right)\quad\left[\textrm{for all }i\in\configspace,\,t\in\setoftimes\right],\label{eq:DilatedProbabilityDistributionMarginalizationCondition}
\end{equation}
 as in \eqref{eq:StandaloneProbabilityFirstSubsystemFromMaginalization}.
\item Let the algebra $\tilde{\rvalgebra}$ of random variables be the set
of all maps of the form $\mapping{\tilde{A}}{\tilde{\configspace}\cartesianprod\tilde{\setoftimes}}{\mathbb{R}}$.
\end{itemize}
These results establish that $\parens{\tilde{\configspace},\tilde{\setoftimes},\tilde{\setoftimes}_{0},\tilde{\stochasticmatrix},\tilde{p},\tilde{\rvalgebra}}$
is a composite unistochastic process, and that the original indivisible
stochastic process $\left(\configspace,\setoftimes,\setoftimes_{0},\stochasticmatrix,p,\rvalgebra\right)$
can be regarded as one of its subsystems. This conclusion completes
the proof of the stochastic-quantum theorem \eqref{eq:StochasticQuantumTheorem}.
QED

\subsection{The Corresponding Quantum System\label{subsec:The-Corresponding-Quantum-System}}

Corresponding to the dilated unistochastic process $\parens{\tilde{\configspace},\tilde{\setoftimes},\tilde{\setoftimes}_{0},\tilde{\stochasticmatrix},\tilde{p},\tilde{\rvalgebra}}$
is a quantum system based on a Hilbert space $\tilde{\hilbspace}$
of dimension $\tilde{N}\leq N^{3}$, with linear-unitary time evolution
encoded in the unitary time-evolution operator $\tilde{\timeevop}\left(t\from0\right)$.

Unistochastic matrices are not generally orthostochastic, meaning
that they are not guaranteed to be expressible in terms of real-orthogonal
matrices. As a consequence, if one is given an indivisible stochastic
process whose $N\times N$ transition matrix $\stochasticmatrix\left(t\from0\right)$
is \emph{already} unistochastic, then there is no guarantee that the
corresponding $N\times N$ unitary time-evolution operator $\timeevop\left(t\from0\right)$
can be assumed to be a real-orthogonal matrix. The stochastic-quantum
correspondence therefore implies that in order to provide Hilbert-space
representations for the most general kinds of indivisible stochastic
processes, the complex numbers $\mathbb{C}$ will be an important
feature of quantum theory.

Whether in that case or more generally, of course, one is always free
to start with an $N\times N$ time-evolution operator $\dynop\left(t\from0\right)$
in \eqref{eq:StochasticMatrixEntryFromAbsValSqTimeEvOp} whose individual
entries are all real. With that choice, the $\tilde{N}\times\tilde{N}$
unitary time-evolution operator $\tilde{\timeevop}\left(t\from0\right)$
obtained from the Stinespring dilation theorem will likewise be real,
and will therefore be an orthogonal matrix. 

However, it is important to keep in mind that from the point of view
of the stochastic-quantum correspondence, the Hilbert-space representations
of a given indivisible stochastic process are convenient fictions,
and so one is entirely free to assume that they involve the complex
numbers anyway. Assuming that a given choice of Hilbert-space representation
is defined over the complex numbers, rather than merely over the real
numbers, allows one to take advantage of the spectral theorem, eigenvectors
of the time-evolution operator $\tilde{\timeevop}\left(t\from0\right)$,
and anti-unitary operators. Assuming appropriate smoothness conditions
in time, one can further make use of a self-adjoint Hamiltonian $\tilde{H}\left(t\right)$
with real-valued energy eigenvalues, as well as the Schrödinger equation.
To the extent that these mathematical constructs are often taken by
textbooks to be indisputable features of quantum systems, the complex
numbers become an unavoidable part of quantum theory.\footnote{Even if one assumes that a given choice of Hilbert-space representation
is defined over the complex numbers, one can always double the dimension
of the Hilbert space from $N$ to $2N$ and represent the imaginary
unit $i\defeq\sqrt{-1}$ by the $2\times2$ real matrix $\left(\begin{smallmatrix}0 & -1\\
1 & 0
\end{smallmatrix}\right)$, in which case all $N\times N$ unitary matrices become $2N\times2N$
real orthogonal matrices (Myrheim 1999)\nocite{Myrheim:1999qmoarhs}.
Interestingly, one can then also represent the complex-conjugation
operation $K$ needed for anti-unitary operators as a $2\times2$
real matrix $\left(\begin{smallmatrix}0 & 1\\
1 & 0
\end{smallmatrix}\right)$, which happens to coincide with the first Pauli sigma matrix $\sigma_{x}$
(Stueckelberg 1960)\nocite{Stueckelberg:1960qtirhs}. However, this
approach still ultimately preserves the algebraic structure of the
complex numbers in the Hilbert space, in the sense that they will
correspond to a subalgebra of matrices on the overall Hilbert space
that commute with the Hamiltonian and all the system's observables.
Hence, this approach does not truly eliminate the complex numbers
from the Hilbert-space formalism of quantum theory. Moreover, using
this \textquoteleft real\textquoteright{} representation means that
the Hilbert spaces of composite systems will not factorize as neatly
into Hilbert spaces for their constituent subsystems.}

\section{Discussion and Future Work\label{sec:Discussion-and-Future-Work}}

The unitarily evolving quantum system that lies on the other side
of the stochastic-quantum correspondence is not limited to a \emph{commutative}
algebra of observables represented by operators that are diagonal
in the configuration basis.

Indeed, as explained in other work (Barandes 2025)\nocite{Barandes:2025tsqc},
one can model the quantum measurement process of an observable represented
by an \emph{arbitrary} self-adjoint operator in terms of a unistochastic
process that contains a subject system, a measuring device, and an
environment, all as explicitly defined subsystems. By taking the measuring
device's allowed configurations to correspond to definite readings
of outcomes, and by taking the transition matrix for the overall unistochastic
process to be based on precisely the type of unitary time-evolution
operator employed in standard textbook treatments of the measurement
process, one inevitably finds that the measuring device ends up in
one of its measurement-reading configurations with a stochastic probability
given by the general form of the Born rule. Hence, in principle, one
has access to a quantum system's entire \emph{noncommutative} algebra
of observables.

In essence, the stochastic-quantum correspondence therefore suggests
that every quantum system may ultimately be an indivisible stochastic
process in disguise. From this standpoint, the usual Hilbert-space
formulation is then just a convenient mathematical layer on top of
a system evolving stochastically along some trajectory through a configuration
space in a highly non-Markovian way. 

With the stochastic-quantum correspondence in hand, one can refer
other exotic features of quantum systems back to their associated
indivisible stochastic processes to give those features a more physically
transparent interpretation. For example, interference and entanglement
can be understood as artifacts of the generic indivisibility of the
stochastic dynamics. Moreover, from the perspective of this stochastic-quantum
correspondence, what gives quantum computers their possible advantage
over classical computers is their access to dynamical laws that are
indivisible, and therefore extremely non-Markovian. 

This overall approach to quantum foundations therefore sheds new light
on some of the strangest features of quantum theory, in addition to
suggesting novel applications of quantum computers. This approach
might even provide a helpful stepping stone for the development of
self-consistent generalizations of quantum theory itself.

\section*{Acknowledgments}

The author would like to thank Scott Aaronson, Emily Adlam, David
Baker, Craig Callender, Iris Cong, Jenann Ismael, David Kagan, Alex
Meehan, Logan McCarty, Wayne Myrvold,  Jill North, and Noel Swanson
for helpful conversations. This research was supported by the American
Institute of Mathematics.

\bibliographystyle{1_home_jacob_Documents_Work_My_Papers_2023-Stoc___ses_and_Quantum_Theory_custom-abbrvunsrturl}
\bibliography{0_home_jacob_Documents_Work_My_Papers_Bibliography_Global-Bibliography}

\end{document}